\documentclass[twocolumn]{aastex631}

\usepackage{graphicx}
\usepackage[caption=false]{subfig}	
\usepackage{longtable}
\usepackage{latexsym}
\usepackage{amsmath}
\usepackage{comment}
\usepackage{scalerel}
\usepackage{xcolor}

\newcommand{\wise}{\textit{WISE}}
\newcommand{\WISE}{\textit{WISE}}

\newcommand{\MIR}{mid-IR}
\newcommand{\NIR}{near-IR}

\newcommand{\Mstar}{M$_{\scaleto{\star}{5pt}}$}
\newcommand{\Msun}{M$_\odot$}
\renewcommand{\micron}{\,$\mu$m}
\newcommand{\Lsun}{L$_\odot$}

\newcommand{\arcs}{\arcsec }

\newcommand{\Lw}{L$_{\rm W1}$}
\newcommand{\Deg}{$^{\circ}$}

\newcommand{\WIWII}{$\rm W1{-}\rm W2$}
\newcommand{\WIIWIII}{$\rm W2{-}\rm W3$}
\newcommand{\WIWIII}{$\rm W1{-}\rm W3$}
\newcommand{\WIIIWIIII}{$\rm W3{-}\rm W4$}

\newcommand{\Y}{$\Upsilon_\bigstar$}
\newcommand{\Yw}{$\Upsilon_\bigstar^{3.4\mu m}$}
\newcommand{\Yi}{$\Upsilon_\bigstar^{3.6\mu m}$}

\shorttitle{Stellar Mass with \WISE}
\shortauthors{Jarrett, Cluver et al.}

\graphicspath{{figures/}}

\begin{document}

\title{A New \WISE\ Calibration of Stellar Mass
\footnote{Released on January, 2023}}

\author[0000-0002-4939-734X]{T.H. Jarrett}
\affiliation{Department of Astronomy, University of Cape Town, Rondebosch,
 South Africa}
 \affiliation{Western Sydney University, Locked Bag 1797, Penrith South DC, NSW 1797, Australia}
 
\author[0000-0002-9871-6490]{M.E. Cluver}
\affiliation{Centre for Astrophysics and Supercomputing,
Swinburne University of Technology,
John Street, Hawthorn, 3122, Australia}
\affiliation{Department of Physics and Astronomy,
University of the Western Cape,
Robert Sobukwe Road, Bellville, 7535, South Africa}

\author[0000-0002-5522-9107]{Edward N. Taylor}
\affiliation{Centre for Astrophysics and Supercomputing,
Swinburne University of Technology,
John Street, Hawthorn, 3122, Australia}

\author[0000-0003-4169-9738]{Sabine Bellstedt}
\affiliation{ICRAR, The University of Western Australia, 35 Stirling Highway, Crawley WA 6009, Australia}

\author[0000-0003-0429-3579]{A. S. G. Robotham}
\affiliation{ICRAR, The University of Western Australia, 35 Stirling Highway, Crawley WA 6009, Australia}

\author[0000-0001-8459-4034]{H. F. M. Yao}
\affiliation{Department of Physics and Astronomy, University of the Western Cape, Robert Sobukwe Road, Bellville, 7535, Republic of South Africa}


\begin{abstract}

We derive new empirical scaling relations between \WISE\ mid-infrared galaxy photometry and 
well-determined stellar masses from SED modeling of a suite of optical-infrared photometry provided by the DR4 Catalogue of the GAMA-KiDS-VIKING survey of the southern G23 field.  The mid-infrared source extraction and characterization are drawn from the \WISE\ Extended Source Catalogue (WXSC) and the archival ALLWISE catalog, combining both resolved and compact galaxies in the G23 sample to a redshift of 0.15.
Three scaling relations are derived:  W1\,3.4\micron\ luminosity versus stellar mass, and \WISE\ \WIWII, \WIWIII\ colors versus mass-to-light ratio (sensitive to a variety of galaxy types from passive to star-forming).   For each galaxy in the sample, we then derive the combined
stellar mass from these scaling relations, producing \Mstar\ estimates with better than $\sim$25-30\% accuracy for galaxies with $>$10$^{9}$\Msun\ and 
 $<$40 - 50\% for lower luminosity dwarf galaxies.   We also provide simple prescriptions for rest-frame corrections and estimating stellar masses using only the W1 flux  and the \WIWII\ color, making stellar masses more accessible to users of the \WISE\ data. 
Given a redshift or distance, these new scaling relations will enable stellar mass estimates for any galaxy in the sky detected by \WISE\ with high fidelity across a range of mass-to-light.

\end{abstract}



\section{Introduction} \label{sec:intro}

Studies of galaxy evolution rely upon accurate estimation of physical properties, most notably the mass metrics such as halo, gas, and stellar mass.  
Since the advent of near-infrared technology, wide-field surveys, and more recently space-based infrared telescopes, the stellar mass (\Mstar) can be effectively estimated from the near-infrared luminosity and a (typically, single) mass-to-light (\Y). The fundamental assumption is that the dominant stellar mass component of a typical galaxy arises from the evolved population of low mass (e.g., Solar-like) stars that are best traced in the near-infrared 1-2\micron\ window where this emission peaks, or at slightly longer wavelengths in the mid-infrared ($<$5\micron), where space-based telescopes provide the additional advantage of a stable PSF and low surface brightnesses.

Galaxies are of course comprised of a variety of stars with differing ages, masses, histories, dynamics, and metallicity.  Attempts to better estimate the \Mstar\ and \Y\ distribution in galaxies typically use sophisticated SED modeling across a wide energy spectrum, but this can be very expensive and limited in areal survey coverage.  A hybrid solution is to create a set of simple scaling functions that connect or ``bootstrap" basic infrared photometric measurements (e.g., color) to well-determined (e.g., SED modeling) stellar masses, thus extending the \Mstar\ calibration to much larger surveyed areas without the benefit of extensive multi-band photometry.  Indeed, a single color is often more than adequate to estimate the global \Y\ for a galaxy \citep[e.g.,][]{Bell&deJong+2001,taylor+2011,robotham+2020,Li-Leja22}.


For near-infrared (NIR) photometric bands the mass-to-light is 
optimally near unity 
\citep[within 2$\times$; cf.][]{Bell+2003,taylor+2011},
while the effects of extinction are far less important compared to the shorter wave bands.  This was a key selling point of the 2MASS 1-2\micron\ all-sky survey back in the 1990s:  the near-infrared window traces the dominant stellar backbone of galaxies and is less affected by dust extinction, and hence an ideal probe of stellar mass.  By today's standards, the 1\,mJy flux limits of 2MASS are relatively shallow, limiting studies to very local volumes \citep[e.g., redshifts less than 0.05 for the 2MRS redshift survey;][]{huchra+2012}.  More recent infrared surveys 
\citep[e.g., the VISTA Hemisphere Survey or VHS;][]{mcmahon+2013}.
are several orders of magnitude deeper and hence enable studies to extend further in distance, while also revealing the lower surface brightness emission in nearby galaxies that may be completely hidden in the shallow surveys.  

One such deep survey, covering the near to the mid-infrared window, comes from the Wide-field Infrared Survey Explorer (\wise) launched in 2009, that fully surveyed the entire sky in four bands from 3 to 23\micron\ (Wright+2010).  The mission was designed to be sensitive to both stellar emission and the interstellar medium of galaxies.  Two near-infrared bands, W1 (3.4\micron) and W2 (4.6\micron, capture the Rayleigh-Jeans (RJ) tail of the stellar continuum that arises from evolved stars, while the mid-infrared bands at W3 (12\micron) -- broad enough to capture both atomic and molecular lines and continuum emission -- and W4 (23\micron) are sensitive to small-grain dust continuum that arises from the star formation processes that build new mass as galaxies evolve \citep{Jarrett13}.   

Since the launch of \WISE, many studies have used the \MIR\ to probe the current and past star formation history properties of galaxies.
Similar to the Spitzer-IRAC 3.6\micron\ band in which a number of studies have shown to be an effective metric for the mass-to-light and hence stellar mass
\cite[e.g.][]{eskew+2012,meidt+2014,que+2015},
the \WISE\ W1 3.4\micron\ band luminosity (\Lw) may be used to estimate the global stellar mass using a single mass-to-light (M/L) value.

Studies using IRAC 3.6\micron\ point to a \Mstar/L$_{3.6\mu m}$ (\Yi) with a 
value of 0.4 to 0.6 
\citep{mcgaugh+2012,schombert&mcgaugh+2014,meidt+2014}.
Using a single M/L is only an approximation; galaxies are not monolithic populations but contain stars in different stages of evolution and with a variety of physical properties (e.g., metallicity).
A single M/L is best used with early type or spheroidal galaxies, whose emission tends to be dominated by older evolved stars,  comprising  a more homogeneous class that may be described by simple stellar populations.  Conversely, it is  
less effective with late-type or SF disk galaxies whose composition is a mix of old and young populations with distinct dynamical distribution (e.g., disk, bulge, bars, etc).

To allow for variation in the \Y\  across galaxy types, photometric colors are typically used both globally and internally.  
The assumption is that the range in color (or set of colors) will indicate the dominant populations (e.g., disk vs spheroidal) and map to a mass-to-light that is more appropriate to its aggregate \Mstar.
In this way, studies have shown correlations of color with \Y\ for normal galaxies, including those using the IRAC [3.6\micron]-[4.5\micron] color
\citep[e.g.,][]{meidt+2014,que+2015,ponomareva+2018}
and the \WISE\ \WIWII\ color 
\citep[e.g.][]{Cluver14,norris+2014}.
At these wavelengths, \Y\ dependencies with metallicity is likely small, although there are conflicting studies, see for example \cite{norris+2014} and \cite{Rock+2015}.

In addition to dust absorption mitigation, the great advantage of using these 3-5\micron\ mid-IR bands is superior detection sensitivity (relative to ground NIR observations). But as pointed out by
\cite{meidt+2012},
they are also subject to warm dust continuum emission from star formation and AGN activity, and additionally, the 3.3\micron\ PAH emission line that lies within the IRAC-1 and W1 bands.  These non-stellar mechanisms act as contaminants to the luminosity and colors, and caution is in order when working with extrema galaxies (e.g., infrared-bright galaxies).
See \cite{ponomareva+2018}
for a comprehensive review of the efforts to measure the \Y\ using the 3-5\micron\ bands of Spitzer-IRAC.

\vspace{5pt}

The \WISE\ study from \cite{Cluver14} showed that the \WIWII\ color trends such that higher \Yw\  tracks with bluer (passive, early-types) colors, while redder colors track to lower \Yw, consistent with SF disks and late-type galaxies.   The full range in \Yw\ is from 0.8 to 0.2, respectively from early to late-types. 

More recently, a novel ``star formation history" approach was used to estimate the \MIR\ mass-to-light.  \cite{Salim+2016} and \cite{leroy+2019}
have shown that the \Yw\ trends with the specific star formation rate (sSFR) as determined through \WISE\ photometry and SED fitting, 
 with a \Yw\ ranging from 0.2 and 0.5, corresponding to actively building galaxies and passive galaxies, respectively.  Moreover, combining several methods, \cite{schombert+2022} demonstrated that the baryonic Tully-Fisher Relation is improved if the \Y\ range is determined using a wide suite of optical-infrared bands, in conjunction with stellar structure (bulge vs disk) modeling, with the expected behavior of lower values for disk populations (\Y\ ranging from 0.4 to 0.8) and roughly constant, large-values (0.8-0.9) for bulge-dominated systems.  
 
 We see that in a range of methods from single M/L, to colors, SF information, and population-kinematic structure,  both the aggregate and local mass density of a galaxy may be estimated from optical, \NIR, \MIR\ (and combinations thereof) imaging surveys.

In this study, our goal is to update and improve upon the early \WISE\ work by \cite{Jarrett13} and \cite{Cluver14} 
with a new set of scaling relations that use our latest galaxy measurements with \WISE\ (see below), coupled to a new and rigorous stellar mass catalog derived from SED modeling of a complete set of optical-infrared photometry and spectroscopic redshifts from the GAMA-KIDS-VIKING DR4 Catalog of the G23 field in the southern hemisphere 
\citep{driver+22}.

Since the launch of \WISE\ in 2009, our team has been building a resolved galaxy database based on \WISE\ imaging and source extraction called the \WISE\ Extended Source Catalogue (WXSC).
The WXSC improves upon both the imaging and source measurements, compared to the published ALLWISE catalogs that were optimized for point sources, 
through custom mosaic construction 
\citep{jarrett+2012}
and characterization that includes careful foreground star (and background galaxy) removal, local background estimation, and a full suite of size/orientation, photometry, surface brightness, and radial profile measurements
\citep[see][]{Jarrett13,jarrett+2019}

In this new study, we deploy the ``total" fluxes (details next section) to derive the W1 3.4\micron\ luminosity, which will be the fundamental scaling tracer of stellar mass.   However, in order to be sensitive to \Mstar/\Lw\  differences from one galaxy to the next, we also derive scaling relations for the colors that \WISE\ provides, namely \WIWII\ and \WIWIII, which are both sensitive to population differences and SF activity \citep{wright+2010,Jarrett13, Cluver14}.  

In this work we, therefore, construct three stellar-mass scaling relations based on the luminosity and colors that can be accessed using \WISE, harnessing the fidelity of the WXSC to provide the necessary photometric accuracy.   We find that our new \WISE-derived stellar masses are accurate to $\simeq$\,25-30\%, or 0.1-0.2 in dex for Log \Mstar.  
Since \WISE\ is a whole-sky survey, and consequently is an effective means to estimate stellar masses for galaxies anywhere in the sky, it is notably powerful in this day and age of wide-field radio to X-ray surveys, including the southern hemisphere next-generation SKA (radio) and LSST (optical) missions. We note that one can always derive more accurate stellar masses using additional photometry, spectroscopy, kinematic and sophisticated SED modeling, if high-quality data is available.  Yet \WISE\ provides a unique data set that is both uniform in properties, sensitive to the dominant mass populations of most galaxies and covers the entire sky.  Our aim is to produce uniform and consistent stellar masses (and SFRs) for all nearby galaxies and thus provide a point of entry to galaxy evolution studies that rely upon stellar mass and star formation activity metrics.

The paper is organized using the following sections:  (2) data and measurements, including the \Mstar\ catalog and WXSC source characterization, (3) procedure for how the \Y\ scaling relations are constructed and used to derive a final \Mstar\ per galaxy, (4) the primary results that including the scaling relations for 3.4\micron\ luminosity, \WIWII\, and \WIWIII\ colors, (5) simple pragmatic methods computing your own masses,  (6) implications of these new relations and (7) summary.

\vspace{10pt}
Throughout this paper, we frequently interchange \Mstar\ to mean M$_{stellar}$, the global stellar mass for individual galaxies.  M/L refers to the mass-to-light ratio (sometimes denoted using \Y).  The adopted cosmology is H$_0$ = 70\,km s$^{-1}$ Mpc$^{-1}$, $\Omega_{M}$ = 0.3 and $\Omega_{\Lambda}$= 0.7.  All magnitudes are in the Vega system ($\textit{WISE}$ photometric calibration described in \cite{Jarrett2011}.


\section{Data} \label{sec:data}

Our aim is to construct scaling relations between \WISE\ mid-IR galaxy properties (\Lw\ luminosity and colors) and `fiducial' stellar masses that have been recently constructed from the GAMA redshift survey.  In this section, we describe the GAMA measurements and stellar masses, and the \WISE\ source characterization measurements.  For \WISE, we use both the WXSC measurements and those coming from the ALLWISE catalog to include both resolved and compact sources that have been targeted by the GAMA survey.  The redshift range is restricted to less than 0.15 to minimize systematic effects from observational biases and errors that grow with (1+z) (e.g., the k-correction), while also providing a large enough sample to cover all galaxy types, \Y, and stellar mass.


\subsection{\Mstar\ Sample}

The primary \Mstar\ sample is drawn from the Galaxy and Mass Assembly (GAMA) Survey 
\citep{driver+2011},
a spectroscopic survey highly complete down to 19.8 mag in the r-band in its three equatorial bands and to 19.2 mag in $i$-band in the G23 field
\cite[see][]{liske+2015,baldry+2018}.
For this \Mstar\ study, we utilize the GAMA-KiDS-VIKING DR4 catalog
\citep{driver+22}
of the
G23 field, located between right ascension 339\Deg\ to 351\Deg\ and declination from -35\Deg\ to -30\Deg and 
covering 50\,deg$^2$ 
in area.  A total of $\sim$47,000 sources have high-quality redshifts in the G23 field.

The catalog also includes a full suite of multi-band ProFound-derived photometry, notably the optical-infrared (ugrizjhk) bands, as detailed in 
\cite{bellstedt+2020}.
From this rich spectro-photometric data set, the GAMA team produced high-quality stellar masses\footnote{StellarMassesv24 GAMA catalog} through SED modeling with 
a fixed grid of relatively simple stellar population models 
\citep[based on][]{bc+2003} 
that includes 
exponential star formation histories, uniform metallicity,  \cite{calzetti+2001}
dust law and 
\cite{chabrier+2003} IMF.
The models are constrained by Bayesian minimum variance parameter estimation marginalized over uniform priors in population age, dust opacity, and extinction.  Physical parameters are in the co-moving frame system, with the same cosmology used in this study (H$_0$ = 70\,km s$^{-1}$, etc).  Formal errors are approximately 30\% for the aggregate stellar masses.
More details on this method may be found in 
\cite{driver+22} and \cite{taylor+2011}.

Independent comparison comes from a study 
using the same data with a more sophisticated ProSpect modeling treatment 
\citep[see][]{robotham+2020, bellstedt+2020b}
that reveals a 0.1 dex offset with 0.11 dex scatter, but essentially 
no differential systematics as a function of color
\citep[see Sec 5 and Fig 34 of][]{robotham+2020}.  Given that ProSpect retrieves more massive values (by 0.1 $\pm$11\% systematic) compared to color-based masses that this current \WISE\ study uses for its scaling relations, this should be noted for the resulting \WISE\ stellar masses.
Nevertheless, the stellar mass  
is surprisingly robust against the well-known degeneracies between age, metallicity, and dust;   
and indeed, using a single optical color like  (g-i) gives nearly as accurate a mass
\citep{Bell&deJong+2001,taylor+2011,robotham+2020}.

Hence we believe these GAMA-G23 stellar masses are a reliable and accurate calibration and will serve well as the ``fiducial" to scale with the \WISE\ mid-IR luminosities and colors.  Not only will this significantly improve the scaling relations since the 
\cite{Cluver14} work (where masses relied on a scaling factor to convert aperture-derived values to total values),
but also places the \WISE\ ``stellar masses" on the same (co-moving H$_0$ = 70\,km s$^{-1}$ frame and Chabrier IMF) calibration as SDSS, GAMA, DEVILS, WAVES, SAMI, Hector, MAGPI and other major surveys that use SDSS/GAMA data.  In total, over 16000 sources have high-quality global \Mstar\ values for redshifts $<$ 0.15 that will be used in this study.

\subsection{\WISE\ Measurements}

 Resolved galaxies are cataloged in the WXSC (Jarrett et al. 2013, 2019), while compact sources come from the ALLWISE (point-source) catalog \citep{cutri+2012} served by the NASA/IPAC Infrared Science Archive (IRSA). In the rest frame, we will be using aperture-matched colors; \WIWII\ and \WIWIII\ in magnitudes, and the \Lw\ in-band luminosity derived from the W1 3.4\micron\ total flux density.

 
 The WXSC includes full source characterization in the four \WISE\ bands: 3.4, 4.6, 12, and 23\micron, extracted from custom-made mosaics of \WISE\      
single frames using the ICORE software developed
by the \WISE\ team 
\citep[details in][]{jarrett+2012}.   
The frames database also includes images
from follow-up NEOWISE mission observations (W1 and W2 only), and hence creating very deep (typically confusion-limited) mosaics, 1-sigma depths fainter than 23 mag\,arcsec$^{-2}$ (26.7 mag\,arcsec$^{-2}$ in AB mags). 
Native angular resolution is preserved in all four bands.

Source characterization consists of four primary steps for the target galaxy: (1) identify and remove foreground stars and background galaxies, (2) local background estimation, (3) 3-$\sigma$ shape and orientation, (4) aperture and 1-$\sigma$ isophotal photometry, (5) surface brightness and radial profile fitting to derive total fluxes, and other miscellaneous characterization; further details in 
\cite{Jarrett13,jarrett+2019}.
For well-resolved galaxies,
axi-symmetric radial profiles were constructed and fitted with a double-Sersic function, thus attempting to
model the spheroidal and disk population distributions, as well as extrapolate to larger radii to determine total fluxes down to several disk scale lengths.

\begin{figure*}[!t]
\begin{center}
\includegraphics[width=0.497\textwidth]{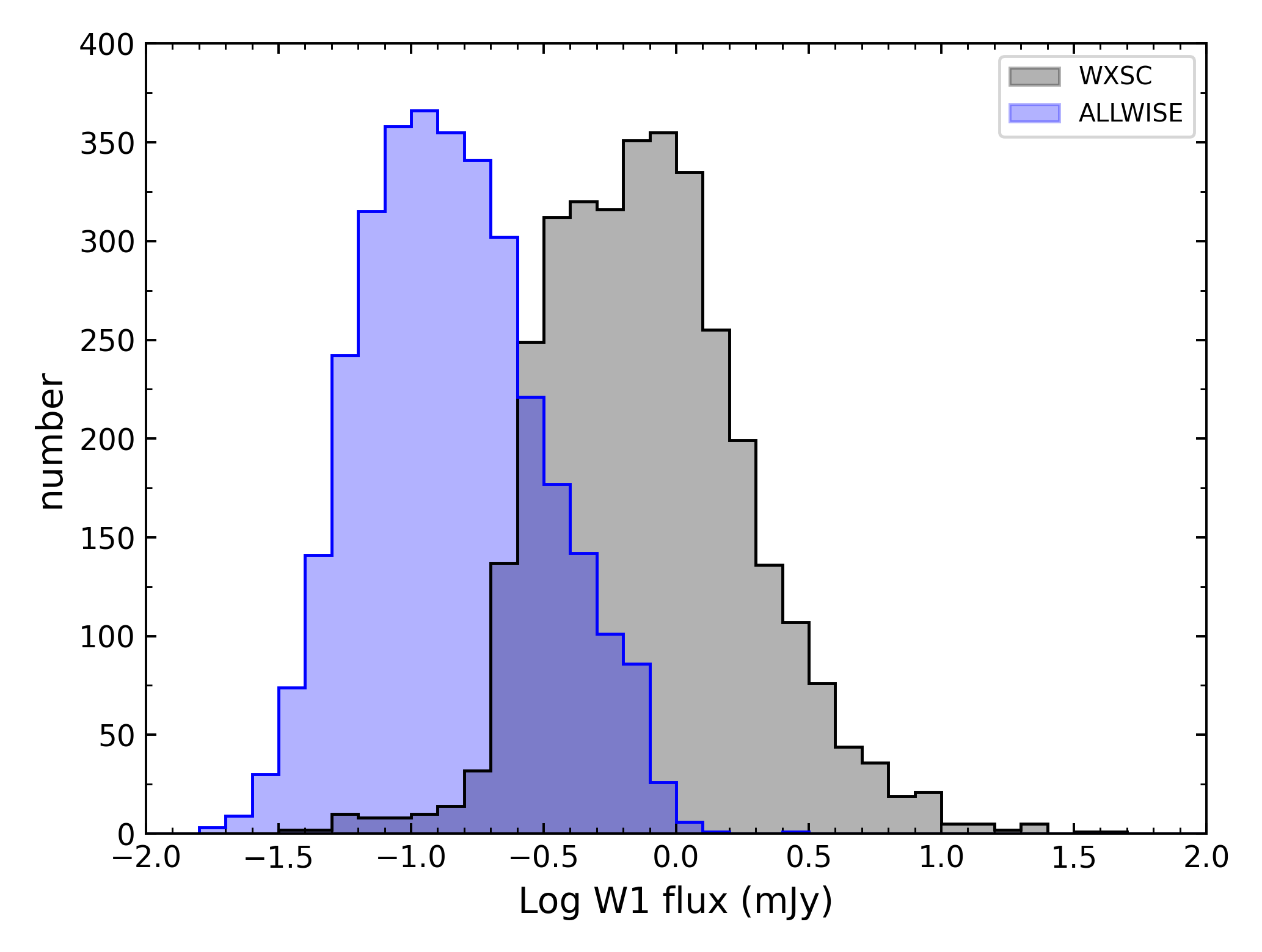}
\includegraphics[width=0.497\textwidth]{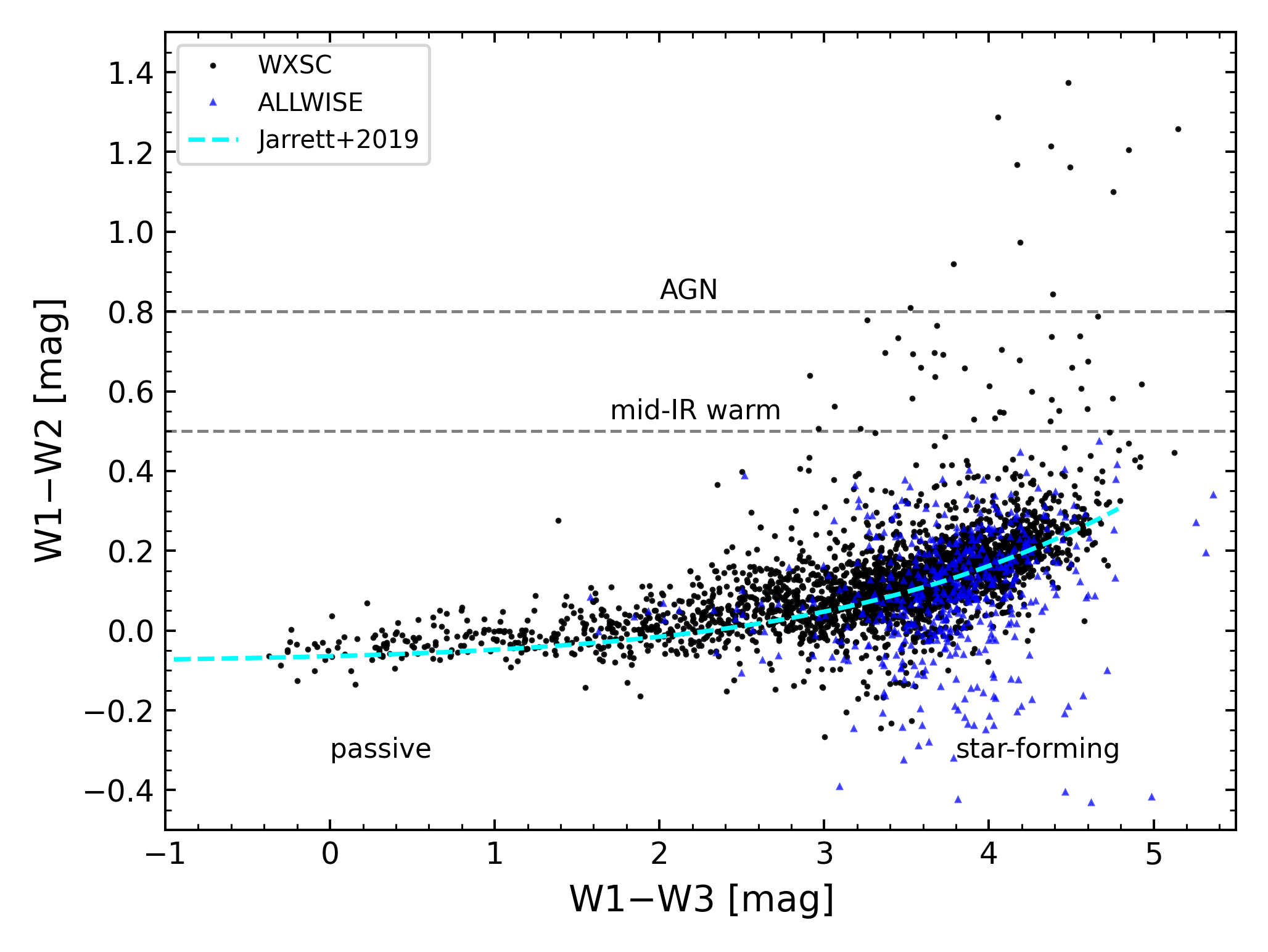}
\vspace{-12pt}
\caption{Mid-IR brightness and color properties of the G23 galaxy sample used to derive the stellar mass scaling relations. 
Resolved sources were measured and cataloged through the WXSC, while fainter (see LEFT panel) compact sources were extracted from the ALLWISE catalog. 
All measurements have been rest-frame corrected (ranging to $z <$ 0.15); see
the Appendix for details and prescriptions for simplifying the process.
For colors (Vega system magnitudes), RIGHT panel, a general trend from left to right is observed, the sequence (cyan dashed line;  
\cite{jarrett+2019}
shows how the colors indicate increasing SF activity from blue to red colors.   A small fraction of the sample has warm \WIWII\ colors indicative of AGN emission significantly contributing to the overall host galaxy emission (and hence eliminated from the analysis).  Table 1 summarizes the number of optical-infrared cross-match detections in these plots.
}
\label{fig:fig1}
\end{center}
\end{figure*}

Typically
the W1 1-$\sigma$ isophote captures 95\% of the total light 
\citep{jarrett+2019},
but some (e.g., low surface brightness) cases
it can be more hidden in the background noise (e.g., 50\% of the light from such a galaxy may be unaccounted).  We, therefore, use total fluxes, when available, to derive the W1 luminosities for the \Mstar\ scaling relation of this study.  
 In the case of compact sources where little emission is detected beyond the 6\arcs\ FWHM beam,  and hence radial profiles are not possible, we instead apply an aperture correction that at the very least accounts for the PSF emission that extends beyond the small, compact aperture. 

 All fluxes and colors are rest-frame corrected, accounting for the (1+z) wavelength stretching,  using a suite of composite SED templates from 
 \citet[]{Brown2014} and Spitzer-SWIRE/GRASIL \cite[]{Polletta2006,Polletta2007,Silva1998}
that range from early-type spheroidals to late-type spirals, Seyferts, and AGNs of differing types. For redshifts $<$0.15, we can expect $<$10-15\% scatter from this k-correction operation based on a recent analysis in \citet[]{yao+2022}.
Further details of the SED template fitting are given in the Appendix, including
prescriptions to estimate the corrections for the flux and \WIWII\ color.
 
  Using the redshift information, transforming to the CMB frame, and using H$_o$=70\,km/s/Mpc cosmology to derive luminosity distances,  
  the rest-frame W1 total fluxes are converted to the in-band
  luminosity, \Lw,
 \begin{equation} \label{eq1}
 L_{\rm W1} [L_\odot] = 10^{-0.4(M-M_{\rm SUN})}
 \end{equation}
where M is the absolute magnitude of the source in W1 (determined from the distance modulus) and 3.24 mag is 
 the W1 in-band solar value 
 \citep{Jarrett13}.  It is important to note that this luminosity is not a spectral luminosity ($\nu$L$_\nu$), but instead is bolometric in nature (normalizing the in-band flux by the solar value in the same band).  This has implications for the k-correction (see Appendix) and means the in-band luminosity is much larger than the spectral luminosity (because the latter is normalized by the full bolometric luminosity of the Sun).    
 
A final note with regard to the 3-5\micron\ light,  we do not attempt to model and correct for any non-stellar emission in the W1 and W2 bands.  It is likely to be negligible, or small relative to the total propagating errors, for most galaxies. For those cases where the mid-infrared is subject to additional AGB dust emission (old stars in the asymptotic giant branch phase), ISM molecular emission (strong PAH emission in starbursts), or AGN-accretion emission (e.g., luminous infrared galaxies), the W1 luminosity (and hence stellar masses) may be over-estimated. We have taken measures (see below when discussing the \WISE\ colors) to eliminate `warm' galaxies (i.e., AGN and powerful starbursts) from the scaling relation analysis to follow; see below.

In the case of compact and point sources, we use the ALLWISE standard aperture magnitudes. 
Nearby galaxies, which tend to be resolved even in the W1 imaging, are not well approximated by PSF profile fitting.
 Compared to the default (PSF-fit) ALLWISE magnitudes,  
 we have determined that the ALLWISE cataloged standard fixed aperture -- small circular aperture, 8.25\arcs\ in radius, corrected to capture the total flux of a point source, is better suited to characterizing nearby, compact galaxies.
As discussed in 
\cite{Cluver14,cluver+2020},
the fixed circular aperture tends to capture more `extended' flux that compact (or nearby galaxies) galaxies will possess.
 Nevertheless, we have found that these standard apertures require further correction to better capture the total extended flux and 
bring them onto a consistent system with the WXSC isophotal magnitudes; 
 hence, we have applied a further correction to the ALLWISE measurements to capture the total flux of the compact extragalactic source.
We can expect 10-20\% uncertainty in the resulting fluxes both from aperture corrections and due to the fainter and hence lower S/N sample relative to the well-resolved galaxies of the WXSC.

Colors are generally determined using matched apertures that are large enough to capture most of the galaxy flux (and hence, representative of the global color).  As detailed in 
\cite{jarrett+2019},
matching W1 and W2 bands is straightforward since they both trace the same emission mechanism (stellar), while matching with W3 (or W4) is not always optimal because of the combined mechanisms of stellar and interstellar medium (gas and dust), as well as the overall (lower) sensitivity of the long wavelength bands of \WISE.    For stellar-dominated galaxies, the W3 and W4 detections tend to be weak or undetermined, resulting in smaller apertures (with the appropriate aperture correction applied, see above), while SF-dominated galaxies have strong W3 detections and aperture matching is possible.  In either case, the \WIWIII\ color is not as stringent as that of \WIWII\ because of the large dynamic range -- 4 to 5 magnitudes vs. 0.5 mag -- so we allow lower S/N detections to create \WIIWIII\ and \WIWIII\ colors.

\vspace{10pt}
Using a typical optical-infrared cross-match cone-radius of 3\arcs\ 
\citep{Cluver14,Jarrett2017},
we positionally matched the G23 \Mstar\ catalog with our rest-wavelength \WISE\ measurements  
to draw the \Mstar-\WISE\ sample of this study.  
We have restricted the sample to $z <$ 0.15 in order to maximize brightness S/N, mitigate luminosity bias and other redshift trends that may skew the scaling relations, minimize rest-frame k-correction uncertainties (see also the Appendix for a discussion on k-corrections) and general completeness and reliability uniformity.
Next, we eliminate suspected AGNs (using \WISE\ colors, see below) and low S/N detections in the W1 (3.4\micron) band.  The resulting match in total with 6671 \WISE\ sources, of which $\sim$51\% are resolved (with S/N $>$ 5 from the WXSC) and 49\% are unresolved or compact (S/N $>$ 3 from ALLWISE).  For the \WIWII\ color, we demand better accuracy for the scaling relation, and hence we set S/N $>$ 7, resulting in a lower number of matches, 3305 and 984 for resolved and unresolved respectively.  Finally, for the \WIWIII\ color, the scaling relation can tolerate a lower sensitivity because of the large dynamic range in the color, and hence the S/N limit is set to 3, resulting in 2692 and 505 matches respectively (note that the detection sensitivity in the W3 12\micron\ band is considerably less than that of W1). A summary of the cross matches between GAMA and \WISE\ is given in Table 1.

In Fig. 1,  we show the rest-frame W1 fluxes and colors 
for the WXSC and ALLWISE measurements that are used in the \Mstar\ calibration (see also Table 1 for a summary of the cross-match statistics).  
Most resolved sources peak around 1 mJy, but are as faint
as 0.1 mJy, while the compact sources peak at 0.1 mJy and reach 
 down to 25\,$\mu$Jy.
In terms of color, 
early-types (spheroidal and passive in SF activity) tend to have blue \WIWIII\ colors (left side of the plot), while dusty SF disks populate the right side (i.e., red colors). 
There is a clear sequence in color-color 
\citep[cyan dashed line, from][]{jarrett+2019}  that shows how the \WIWII\ color 'warms' as SF activity increases, due to the W2 4.6\micron\ band increasingly sensitive to the warm-dust ISM component.  Additional \MIR\ emission may arise from AGN accretion, which tends to further redden the colors, most notably in the \WIWII\ axis, while for dusty (AGN+)starbursts the W1 3.4\micron\ band may be enhanced due to the 3.3\micron\ molecular PAH molecular \citep[typically very weak for normal galaxies, yet has been detected in
active galaxies; cf][]{kim2012}.  

As determined by 
\cite{Jarrett2011} and \cite{stern+2012},
galaxies with \WIWII\ colors larger than 0.8 are clearly dominated by AGN, including QSOs, BLAGNs, and dust-obscured Type-II varieties. However, a number of more recent studies show that lower power AGN may reside in the range from \WIWII\ $>$ 0.4 mag 
\citep[cf][]{assef+2013,yao+2020},
so caution is in order when the color is warmer than the expected sequence value (cyan dashed line in Fig. 1).
Fortunately, the sample has only a few obvious AGNs, and quite a small (statistically insignificant) number of 'warm' AGNs and hybrid-AGN+starbursts (\WIWII\ $>$ 0.4 mag).
We have eliminated all sources with \WIWII\ $>$ 0.4 
from the analysis to follow.   

\vspace{10pt}

\begin{table}[!h]
 \caption{Summary of GAMA-\Mstar\ and \WISE\ Catalogs} \label{table1}
\centering
\small{
\begin{tabular}{l c c c }
\hline  
\multicolumn{3}{c} {GAMA G23 \Mstar\ sources ($z < 0.15$): 16838}\\

         & x-matches  & x-matches  \\ 
 \WISE\    &   (resolved)         &   (unresolved) \\
 \hline
 \Lw\ (SNR $>$ 3) & 3374 & 3297  \\ 
 \WIWII\ (SNR $>$ 7) & 3305 & 984 \\
\WIWIII\ (SNR $>$ 3) & 2692 & 505  \\
            
\end{tabular}
}
\end{table}

\section{Procedure for Deriving \Mstar\ Relations}

In the previous work on deriving the \Mstar-\MIR\ relation, 
\cite{Cluver14}
determined that the color \WIWII\ was correlated with the \Mstar/L ratio, showing a trend in which redder colors had lower mass-to-light ratios. The trend may (broadly) account for population, morphology, and metallicity changes with color, from passive spheroidal galaxies at one extreme to dusty SF disk galaxies at the other. The observed scatter in the \Mstar\ vs. \WIWII\ relation is large ($\sim$50\%), and the derived stellar masses have correspondingly large uncertainties.  Since that work, we have 
significantly improved the WXSC measurements,  both in terms of deriving better isophotal and total fluxes and better matched-aperture colors, as well as improving our rest-frame correction pipeline.   
For this new work, we also investigate the scaling relation between \Mstar\ and the \WIWIII\ color, which is likely a better tracer compared to W2 because of the improved sensitivity in the W1 measurements and the broader range in color.   

Finally, to account for faint sources where colors (such as \WIWII) are poorly determined, we have derived the simplest relation possible, \Mstar-to-\Lw\ in Log-Log space.  This relation has the largest scatter (compared to the color relations) because it does not account for population, metallicity, or other factors that color is tracking.  It does have a big advantage in that it is
the most sensitive band of \WISE:  
the WXSC is fundamentally based on the detection of the W1 3.4\micron\ emission from galaxies.  In effect, the general linear trend between \Mstar\ and \Lw\ indicates that a single \Yw\  (or narrow range thereof) adequately (better than 40-50\%) describes most galaxies.  


The procedure, therefore, is to derive three \Mstar\ (or M/L) relations:  (1)  W1 luminosity,  (2) \WIWII\ color, and (3) \WIWIII\ color.  With these three relations
\footnote{For completeness, we note that surface brightness is also correlated with galaxy property types 
\citep[$cf.$][]{jarrett+2019},
so we investigated the W1 central surface brightness compared to the GAMA stellar masses. However,  we have determined that the (large) resulting scatter does not improve the final derived stellar mass.  We do not show any further results for surface brightness.}
and their corresponding error models, we then derive a weighted average \Mstar\ and corresponding uncertainty. 


\section{\Mstar\ Results}

In this section, we present the scaling relations between the optically determined stellar masses and the primary \WISE\ measurements of the W1 luminosity, \WIWII, and \WIWIII\ colors.  We will use all of this information to create a weighted-average combination that represents the \MIR\ stellar calibration, which may be used for any \WISE\ 3.4\micron\  total flux measurements of extragalactic sources.

\subsection{\Mstar\ vs. 3.4\micron\ Light}

The W1 3.4\micron\ flux is determined for all WISE-cataloged sources (since it is the most sensitive band in \WISE), and hence is the most general of the scaling relations with \Mstar.  
For resolved sources,  W1 detections range from 5 to 17.2 mag (3.1\,Jy to 40\,$\mu$Jy in flux density; see Fig 1a), and corresponding \Lw\ luminosities (Eq. 1) that range from 10$^7$ to 10$^11.8$\,\Lsun.  For compact sources, 
 W1 mags range from 13.5 to 17.9 mag (1.2\,mJy to 21\,$\mu$Jy), thus pushing down to fainter magnitudes and luminosities.
Although most of the resolved sources are more luminous than 10$^9$\,\Lsun, (compact sources stretching 0.5 dex fainter),  
the sample does allow scaling down to the  dwarf galaxy regime, which was a key goal of this study to improve the \Mstar\ estimates for low mass and dwarf galaxy populations.
The scaling results are shown in Fig 2, Log \Mstar\ versus Log \Lw.

\begin{figure}[!h]
\begin{center}
\hspace{-18pt}
\includegraphics[width=0.5\textwidth]{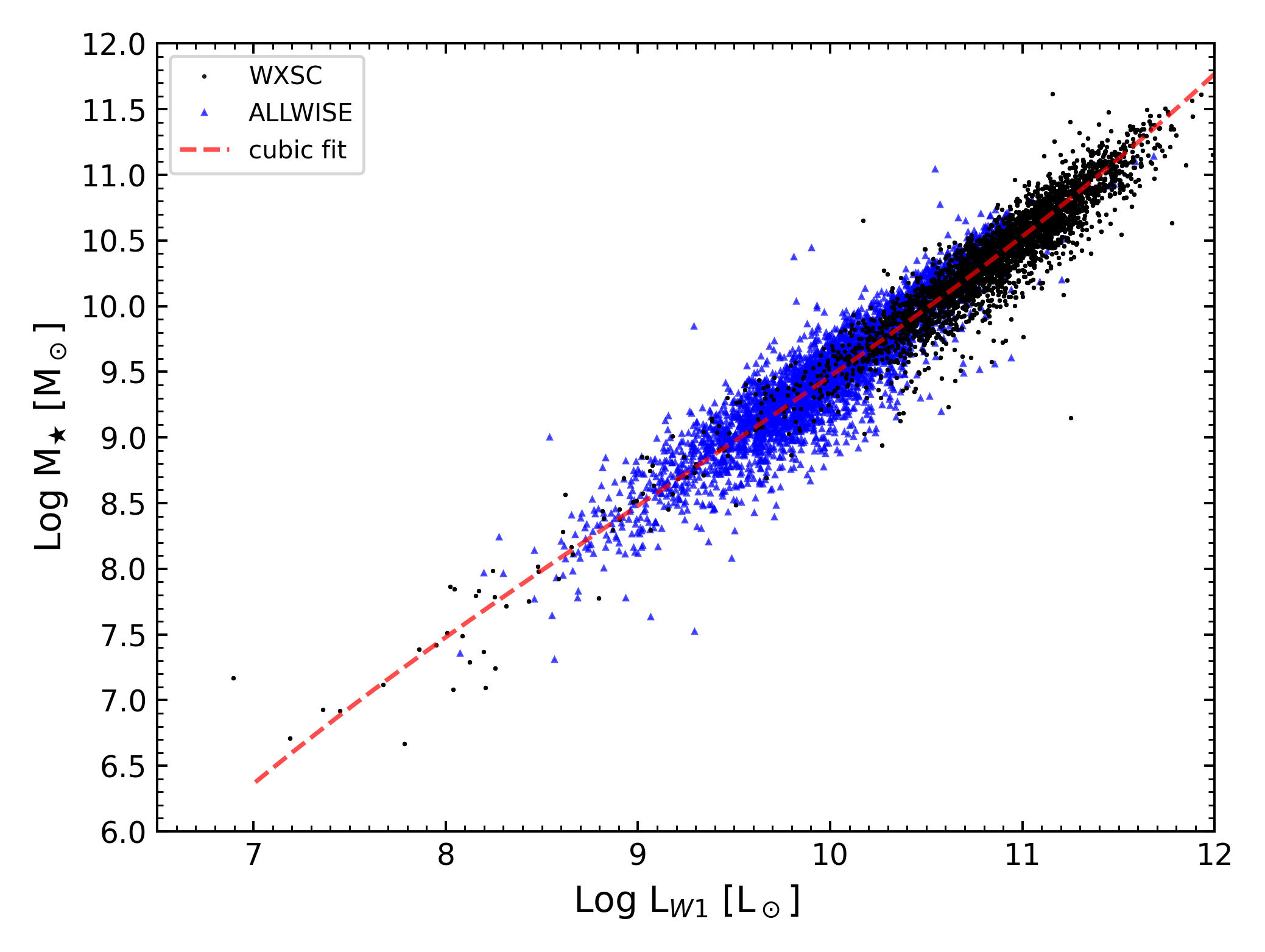}
\caption{W1 scaling relation, comparing the \Mstar-to-light global properties of galaxies as measured in the \WISE\ 3.4\micron\ band. Resolved galaxies from the WXSC are shown in black, whose luminosities are derived from the total integrated flux.   Point sources are shown in blue, whose aperture-corrected measurements are adapted from the ALLWISE catalog.  A 3rd-order polynomial least-squares fit to the distribution (dashed line)  accounts for a small upturn in the \Mstar/\Lw\  at the high luminosity (i.e., massive galaxy) end; see also Eq. 2. Relative to the fit, the vertical scatter is 0.12 dex for masses $>$ 10$^{8.5}$\,\Msun.
}
\vspace{-10pt}
\label{fig:fig2}
\end{center}
\end{figure}

As expected in Fig. 2, luminosity (or ``light") scales with stellar mass, with a range of slope (i.e., mass-to-light) that is highly constrained to values between 0.30 to 0.40 (as shown in the next Section).
The scaling trend is not perfectly linear, there is a small upturn at the luminous end, where the slope is steepening, most notably for the compact sources.  At higher masses, the \Yw\ ratio is steeping, reflecting the population change as older, SF-quenched galaxies dominate at these masses.
Based on the resolved and compact distributions in Fig 2, a least-squares fit using a 3rd-order cubic polynomial covers the range from 7 to 12 in the Log\,\Lw, accordingly:
\begin{equation} \label{eq1}
\begin{split}
Log\,M_{\scaleto{\star}{5pt}}  = A_0  + & A_1 \, (Log\,L_{\rm W1}) + \\
      & A_2 \, (Log\,L_{\rm W1})^2 + A_3 \, (Log\,L_{\rm W1})^3
\end{split}
 \end{equation}
where \Mstar\ is the stellar mass in \Msun,  \Lw\ is the \WISE\ 3.4\micron\ in-band rest-frame luminosity in \Lsun\ (see Eq. 1),  and the A coefficients are -12.62185, 5.00155, -0.43857, 0.01593, respectively.  

The resulting \Yw\ (slope in Fig. 2) has a mean of 0.35\,$\pm$\,0.05.  The vertical scatter in \Mstar\ relative to the quadratic relation is about 0.12 dex (i.e., 32\%), but considerably higher, 0.2 dex, at the low end where the relation is not as well constrained.  This small M/L  range essentially means a single value crudely scales the 3.4\micron\ luminosity to the expected stellar mass.  The small range and random scatter in the trend demonstrate the limits of only using the W1 band to estimate masses, it cannot account for population differences or other properties that may alter the global \Mstar/\Lw.  On the other hand, the cubic polynomial is tracking the change in M/L for the passive galaxies at the high mass end.   
As we shall see below, colors do a better job of tracking \Yw\  differences between galaxies; as such, we will downgrade the W1-\Mstar\ scaling relation when color information is available by lowering its weight by a factor of 5 when weight-combined with the other relations. Nevertheless, with this W1-\Mstar\ scaling relations, there now exists a simple way to derive the stellar mass for faint sources, such as dwarfs, with an accuracy that is 0.1-0.2 dex (30-50\%) in Log \Mstar\ for most galaxy masses.  In the following section, we give an example of how to do simple mass estimation using this relation.

We should point out that these new stellar masses are lower by 0.15 dex than what we derived in 
\cite{Cluver14}
(notably for early-type galaxies, see below for the colors analysis), reflecting the new GAMA \Mstar\ fiducials and improved \WISE\ measurements of the total global flux.
Likewise, the  `single' \Yw\  = 0.35 is considerably lower than the value adopted in the early S4G work with Spitzer IRAC-1 at 3.6\micron\ 
\citep{meidt+2014}.
Conversely, we note that our lower M/L  (and hence, lower \Mstar\ estimates) is more in line with the recent work of 
\cite{ponomareva+2018} and \cite{leroy+2019}.


\subsection{Colors}

\subsubsection{\WIWII}
Applying a S/N limit of 7 to the \WIWII\ colors restricts the sample to 3305 and 984 WXSC and ALLWISE sources, respectively.  The limit was chosen to avoid undue scatter in the relation given the small range or tolerance in color, while also maximizing the useful range as demonstrated in Fig. 1b for normal galaxies.   The resulting scaling with \Yw\  is shown in Fig 3, where we have fit the distribution using a simple linear function that ranges across -0.2 $<$ \WIWII\ $<$ 0.4 mag, and corresponding M/L  range of $\sim$0.2 to 0.6.  

\begin{figure}[!thb]
\begin{center}
\hspace{-18pt}
\includegraphics[width=0.5\textwidth]{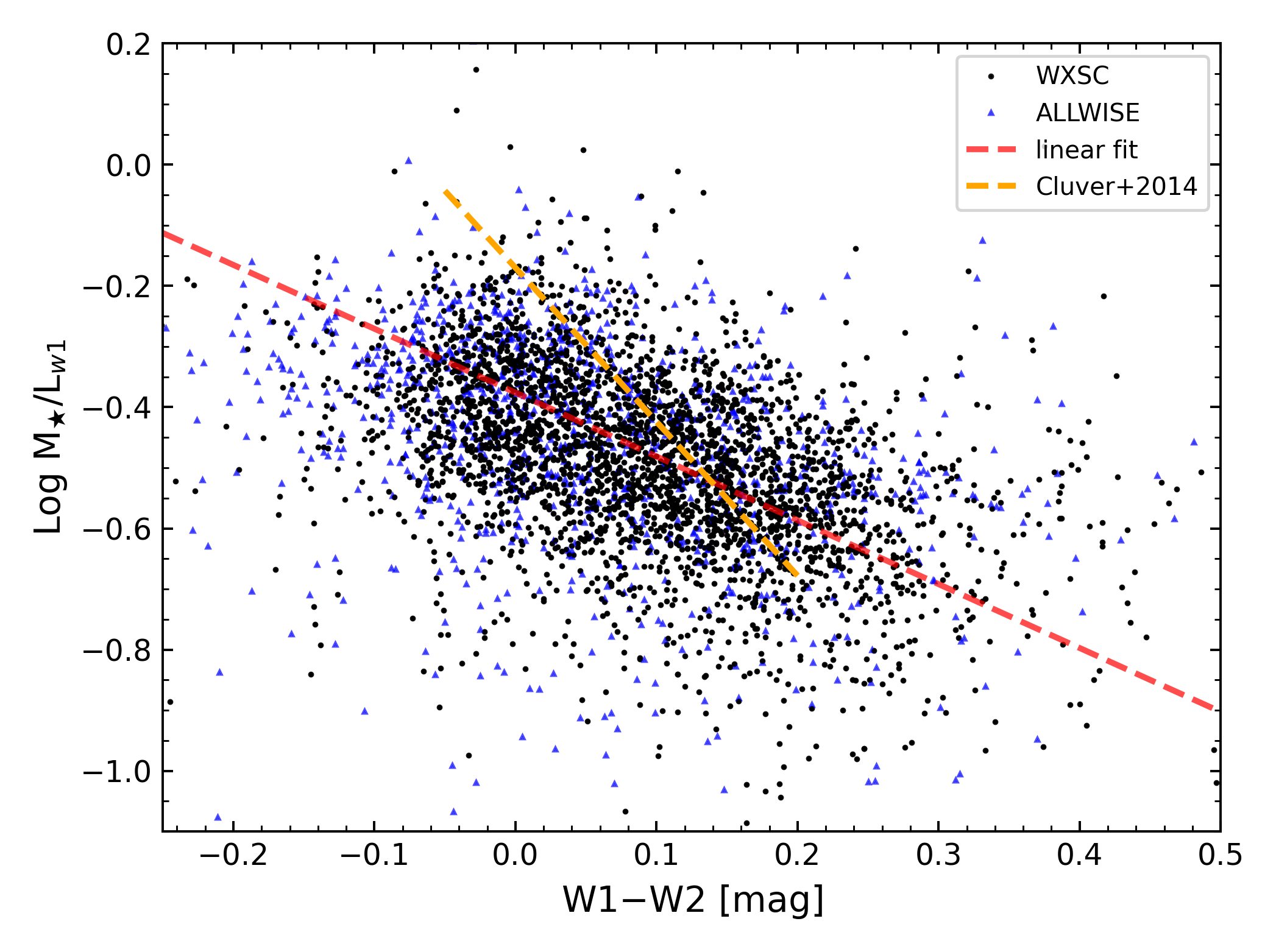}
\caption{Mass-to-light trend with \WIWII\ color (Vega magnitude).  Black and blue points indicate resolved and point-like measurements, respectively. The \WIWII\ color is sensitive to warm dust associated with SF activity (and AGN, if present), see Fig. 1b, and hence the linear trend (red dashed line) from higher to lower \Yw\  follows from early-type (bluer colors) to later-types (redder colors).  Relative to the fit, the vertical scatter is 0.11 dex. 
For comparison, we also show the relation from 
\cite{Cluver14},
indicating previously larger overall \Mstar\ at the high mass (low color value) end.
}
\vspace{-15pt}
\label{fig:fig3}
\end{center}
\end{figure}

The \WIWII\ color shows a wide range in \Yw, trending from high M/L  with blue (stellar-dominated) colors to low M/L  with red (dusty ISM) colors.  Least-square fitting to the distribution, we derive the simple scaling relation:

\begin{equation} \label{eq1}
\text{Log} \Upsilon_\bigstar^{3.4\mu m} = A_0 + A_1 \, C12
 \end{equation}
where C12 = \WIWII\ color is in Vega magnitudes, and the A coefficients are 
(-0.376, -1.053, respectively).  For more extreme colors, the floor/limits should be -0.2 and 0.4 mag, respectively, in the \WIWII\ color, corresponding to \Yw\ range 0.68 to 0.16, respectively.
Compared to the linear least-squares fit, the scatter is about 0.11 dex (in Log\,\Yw) for the entire range.    
In terms of the associated error model, in addition to this vertical scatter,
we have propagated the formal uncertainties in the color and W1 luminosity.

For comparison in Fig. 3, we show the relation from 
\cite{Cluver14}
(orange dashed line), demonstrating that the stellar masses and WISE-GAMA photometry from that study formed a steeper and higher offset M/L  relation.  
Galaxies with bluer colors, typically passive or spheroidal types, will thus have larger masses in the previous color scaling.  For example, at a color of 0.0 mag, the older relation gives a \Yw\  $\sim$ 0.68, while the new relation indicates a lower value of 0.42; hence, in terms of Log \Mstar, the difference is about 0.21 in dex. At the most extreme blue color limit for the 2014 relation, \WIWII\ = -0.05, the difference in Log \Mstar\ is 0.26 dex.
On the other hand, the two relations are in close agreement for warmer colors corresponding to late-type and dusty SF galaxies, \WIWII\ $\sim$0.14 mag;  while for reddest colors, the steeper 2014 relation (which is limited to a color of 0.2 mag) will give lower \Yw\  values:  0.21 versus 0.26 for the new scaling (equivalent to 0.09 dex in the Log \Mstar).


\vspace{10pt}

\subsubsection{\WIWIII}
Next, we considered the ISM-sensitive (i.e., star-formation activity) color, \WIWIII.  Here we relaxed the S/N criteria to 3-$\sigma$ color uncertainty limit to include as many sources as possible.
Accordingly, 2692 resolved (and 505 compact) sources were used in the scaling relation; plotted in Fig 4, showing the Log\,\Yw-[\WIWIII] distribution.

\begin{figure}[!thb]
\begin{center}
\hspace{-18pt}
\includegraphics[width=0.5\textwidth]{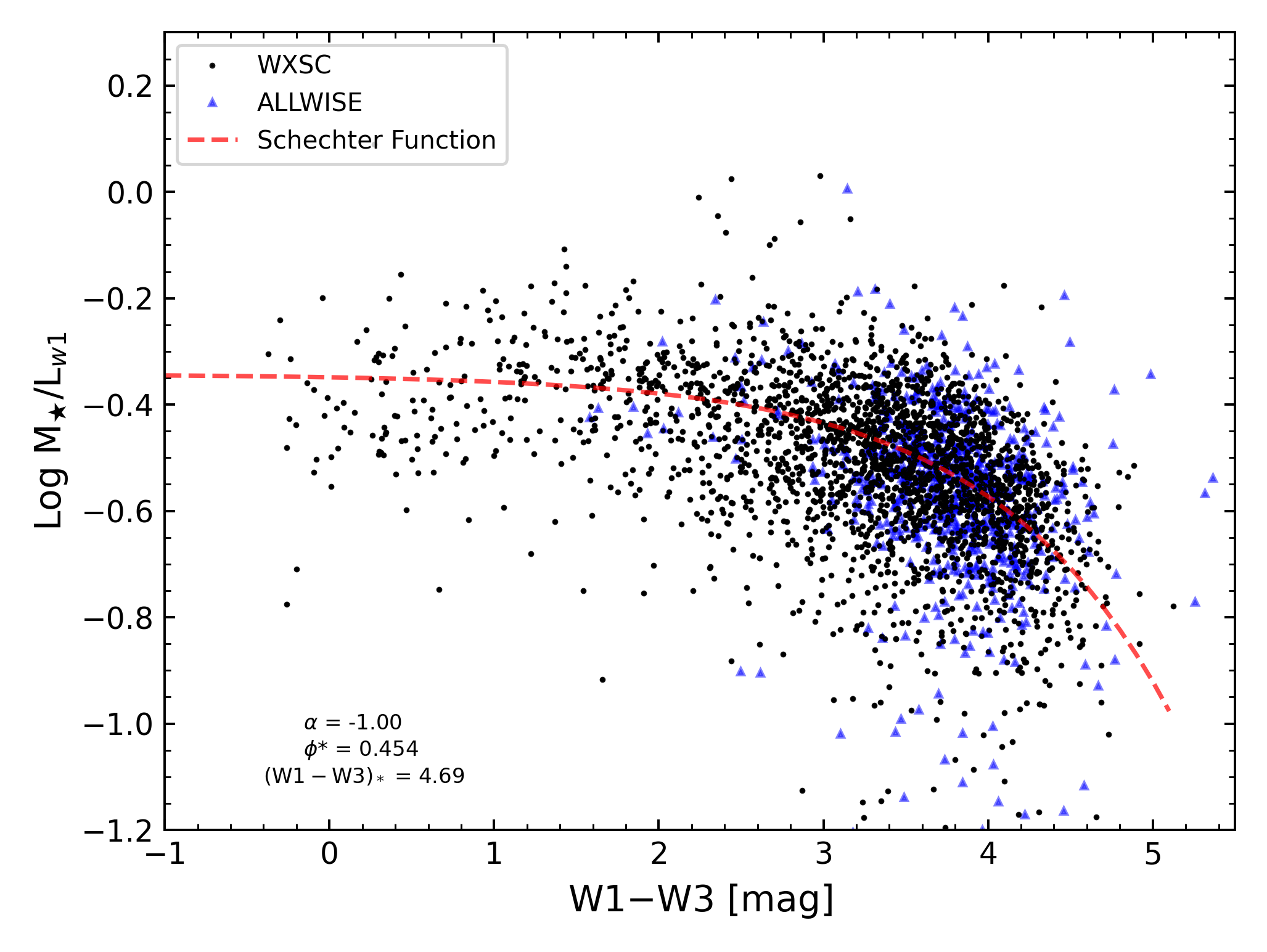}
\caption{Mass-to-light (\Yw) trends with \WIWIII\ color.  This color is notably sensitive to dust content and SF activity (see also Fig. 1b).   As expected, the \Yw\  tends to decrease from blue to red colors (i.e., passive to SF active) but has a non-linear distribution that is suitably modeled by a Schechter Function (red dashed line;  best-fit parameters indicated in the lower left). Relative to the fit, the vertical scatter is 0.09-0.15 dex, increasing to larger values for redder colors.  
}
\vspace{-10pt}
\label{fig:fig4}
\end{center}
\end{figure}

Unlike the \WIWII\ relation, here the \WIWIII\ comparison to the \Yw\  indicates two distinct trends: somewhat flat M/L for blue colors and a steeper trend for red colors.  Specifically at the blue end, 
Fig. 4 shows a flat (single \Yw\ \,$\sim$\,0.45) for a range in color from 0 to 2 mag in the \WIWIII\ color, essentially covering the range that  stellar-dominated galaxies occupy, including early-type and bulge-dominated spirals. Note that this M/L is higher than the overall (all-galaxies) value determined with the simpler \Mstar-\Lw\ relation, consistent with these types of galaxies having a more massive bulge population that is reflected in the \WISE\ color.  For redder colors,  $>$3 mag,   
the distribution in Fig. 4 then rolls off to lower M/L, the vast majority between 0.2 to 0.4, thereafter flooring around 0.15. 

To create a \WIWIII\ scaling relation with \Yw, the two trends may be fit with corresponding linear functions, joining (in discontinuity) at a color of 3.1 mag or so.  It may also fit (somewhat poorly) with a simple exponential function.
However, we also note that the overall      
 Log-Log distribution conveniently resembles a Schechter Function; hence,  we have carried out a least squares fit using a model with variables $\phi_\star$ (vertical scaling factor), $\alpha$ (controls the slope of the flat trend at the blue end), and (\WIWIII)$_\star$ which controls the `knee' bend; accordingly:

\begin{equation} \label{eq1}
\begin{split}
\text{Log} \Upsilon_\bigstar^{3.4\mu m}   = \phi_{\star}\:\gamma^{(\alpha+1)} exp (-\gamma) \\
\gamma  = 10^{[0.4(C13 - C13^\star)]}\\
C13 = W1{-}W3
\end{split}
\end{equation}\\
where we have fit for $\phi_{\star}$, $\alpha$ and color (\WIWIII)$^\star$;  the resulting best fit values are, respectively 0.454, -1.00, 4.690, also indicated in Fig. 4. 
For more extreme colors, the floor/ceiling color limits should be -1.0 and 5.0 mag, respectively, in the \WIWIII\ color, which corresponds to \Yw\  values of 0.45 and 0.12, respectively.
Compared to the best fit function, the vertical scatter is about 0.09-0.15 dex (in Log\,\Yw\ ) for the full range, with increasing uncertainty at the red color extreme where it is poorly constrained.    
As with the \WIWII\ error model, we propagate the vertical scatter, color uncertainty (horizontal scatter), and W1 luminosity uncertainty.

\begin{figure*}[!t]
\begin{center}
\includegraphics[width=0.497\textwidth]{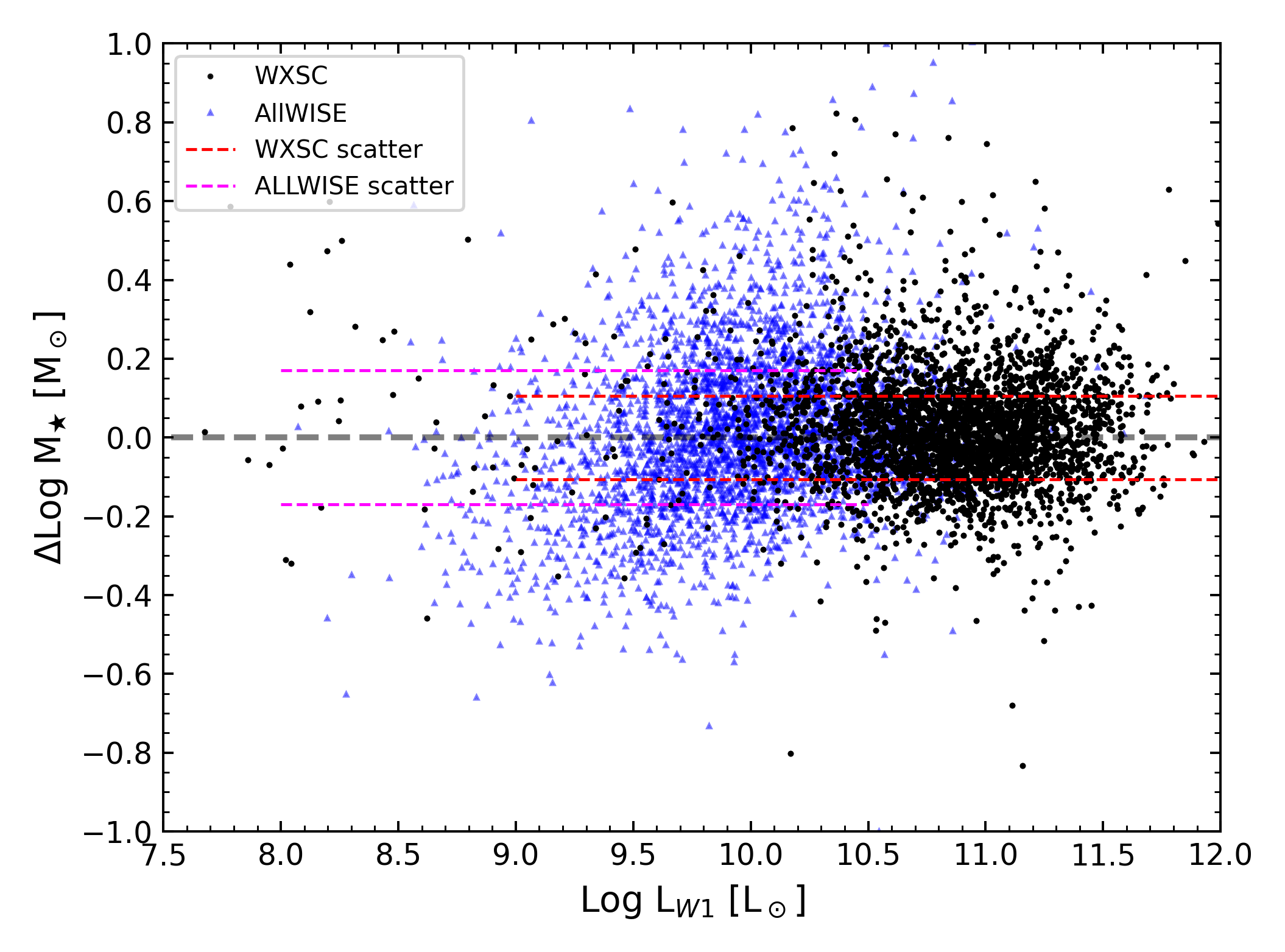}
\includegraphics[width=0.497\textwidth]{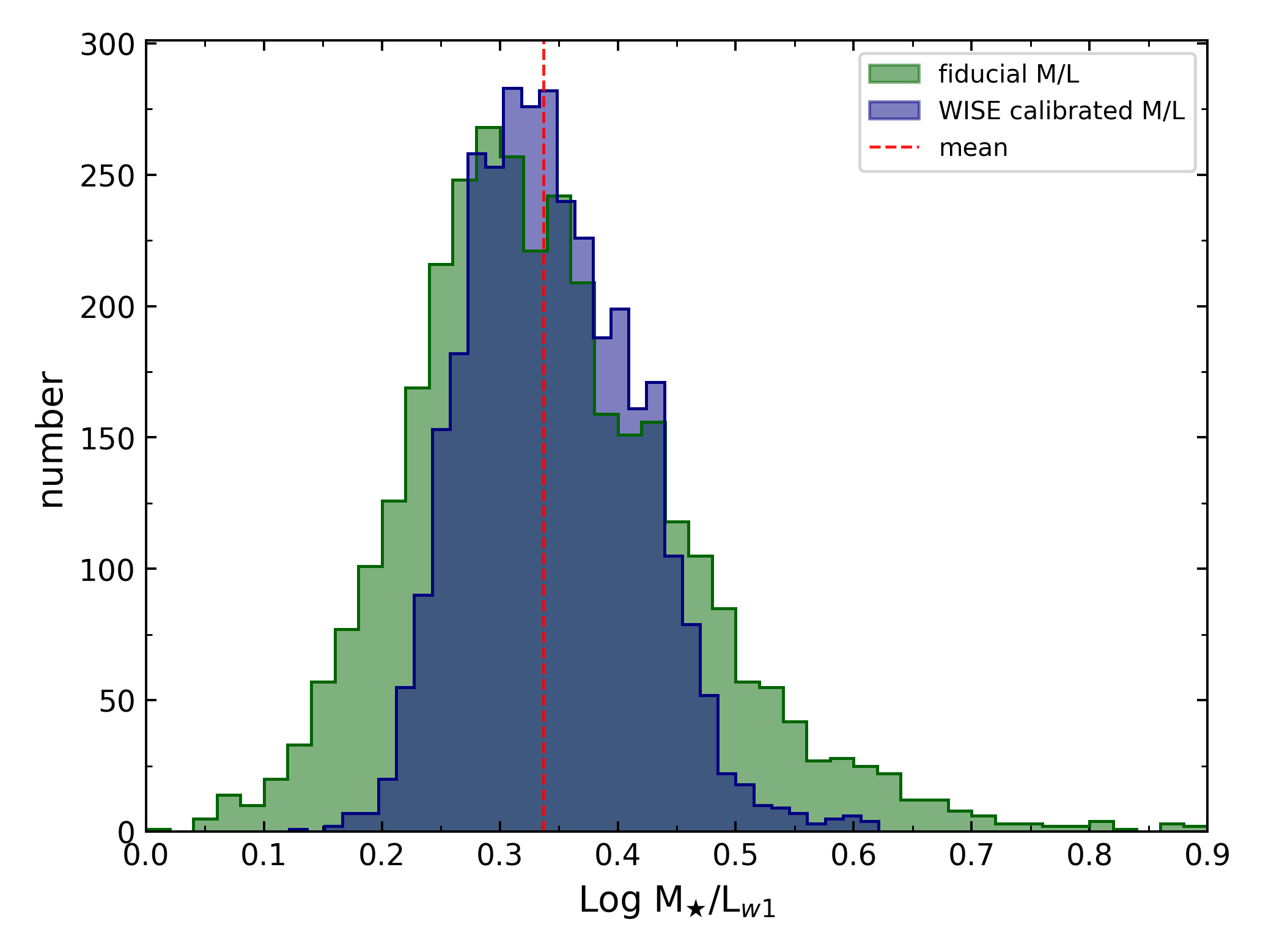}
\caption{\WISE\ \MIR\ stellar masses derived from the combined \WISE\ W1 luminosity and \MIR\ color scaling relations.  (left) shows the newly estimated masses compared to the ``fiducial' masses from which the scaling relations were derived.  
Resolved WXSC (black points) sources have a residual scatter around the zero line of $\sim$0.1-0.12 dex (red dashed line) in the \Lw\ range from 10$^9$ to 10$^{12}$\,\Lsun;  aperture-corrected point sources (blue) have a larger scatter of 0.16-0.17 dex for a luminosity range between 10$^{8.5}$ to 10$^{10.5}$\,\Lsun\ (magenta dashed line).  (right) Corresponding mass-to-light histograms for the complete WXSC distribution (dark blue line), which may be compared with fiducial \Yw\  (green line).  The mean \Yw\  is 0.34, and a full-width half-maximum range 0.23 to 0.44.
}
\vspace{-10pt}
\label{fig:fig5}
\end{center}
\end{figure*}

\vspace{20pt}
\subsection{Combined \Mstar\ Relations:  \WISE\ Stellar Masses}

Combining the estimated stellar masses from the three scaling relations and their corresponding error models,  using an inverse-variance weighting scheme, the \MIR\-based stellar mass is derived.  
The color relations have a smaller total scatter compared to the \Lw\ relation, and hence will have a higher weighting in general.  However as noted earlier, our aim is to cover as wide a M/L range as possible to better model real galaxies, and consequently the \Lw\ relation (Eq. 2) is 
further de-weighted (by a factor of 5) 
in order to emphasize the color relations (Eqs. 3 and 4) which better trace the variation in \Yw\  that is exhibited in the fiducial `truth' set.  The corresponding weighted mean in the global stellar mass is shown in Fig. 5 for the WXSC and ALLWISE sub-samples, where we have compared the \MIR\ masses and M/L with the fiducial counterparts to assess the new scaling relations performance.

The resulting  scatter seen in the Log\,\Mstar\ residual is shown in Fig. 5a (left panel).  For resolved sources (black points), most of the points are in the luminosity range from 10$^9$ to 10$^{12}$\,\Lsun, scatter around the zero residual line with a 0.1 to 0.12 standard deviation, but may also have extremes as deviant as 0.5 in dex.   This empirical result is mostly consistent with the propagated error models which predict most sources will have Log\, \Mstar\ errors between 0.09-0.11 dex (25-30\%).  Since there is some cross-correlation between the three scaling relations (they do share information, notably the W1 flux), we can expect the standard error of the weighted mean to have a slightly lower value than the observed residual scatter.  
 Yet, the  residual
 is still remarkably tight considering that we are only using three pieces of \MIR\ characterization information to derive Log M, all of which are cataloged in the WXSC.  
 
 Remember the formal errors in the GAMA masses is ~0.12 dex (30\%), and hence appear to be the dominant uncertainty in these new \WISE\ scaling relations (as opposed to the photometry).  Moreover, 
 given the relatively flat, zero systematic between the fiducial and the derived, and the 0.1-0.12 dex residual scatter in Log \Mstar, it is reassuring that our mid-infrared bands are not obviously contaminated by non-stellar emission or observational (e.g., redshift) biases.

For compact source from ALLWISE (blue points) the scatter is considerably larger, 0.16-0.17 dex, and some fairly large deviations of $\pm$0.75 dex, but also extending to 
lower luminosities (and hence, lower masses), $\sim$10$^{8.5}$\,\Lsun.  
The error models predict Log\,\Mstar\ uncertainties from 0.12-0.16 dex.
The added variance in the unresolved/compact sources is due to the fact that most are lower S/N and have significantly fewer complete colors ($<$30\%; see Fig. 1a and Table 1), but may also reflect the limitations with estimating ``total" fluxes using ALLWISE aperture-corrected measurements of compact (and likely resolved to some extent) galaxies.


Finally, we consider the resulting \Yw\ , comparing the derived M/L to the fiducial counterpart, shown in Fig. 5b.
The resulting M/L  distribution (black line), has a peak or mode at \Yw\ = 0.34,  a mean value of 0.343, and 
a FWHM width of 0.21 (bearing in mind the distribution is not exactly Gaussian).  Nearly all sources have a M/L between 0.2 to 0.6.  Compared to the fiducial \Yw\ values (i.e., the `truth' set; see green line histogram), which has a mode near 0.35, the fiducial distribution is slightly wider in width (and hence, a larger range in M/L), notably the extreme or wings of the distribution.   
  The slightly narrower range in M/L for the \WISE\ values 
is a reminder that using only limited broad-band information to derive the global stellar mass does have its limitations in tracing the full diversity of M/L that galaxy populations exhibit.
Nevertheless, with these new scaling relations, it will be possible to estimate the Log\, \Mstar\ for any galaxy in the sky with 25-50\% fidelity.

\section{Practical Matters:  Simplified k-corrections and Stellar Mass}

The method for estimating stellar masses presented thus far uses the W1 luminosity and two colors of \WISE.   These are then weight-combined into a mean value based on their excepted variance.  Moreover, the measurements have been rest-frame corrected.  The latter is particularly daunting with the \WISE\ bands (see the Appendix).   In this section, we offer some simple prescriptions for estimating masses that help overcome the data complexities.  Although the resulting masses may not be as reliable as the robust determination presented thus far, they nevertheless will be estimates that suffice for most needs.

\begin{figure*}[!t]
\begin{center}
\includegraphics[width=0.497\textwidth]{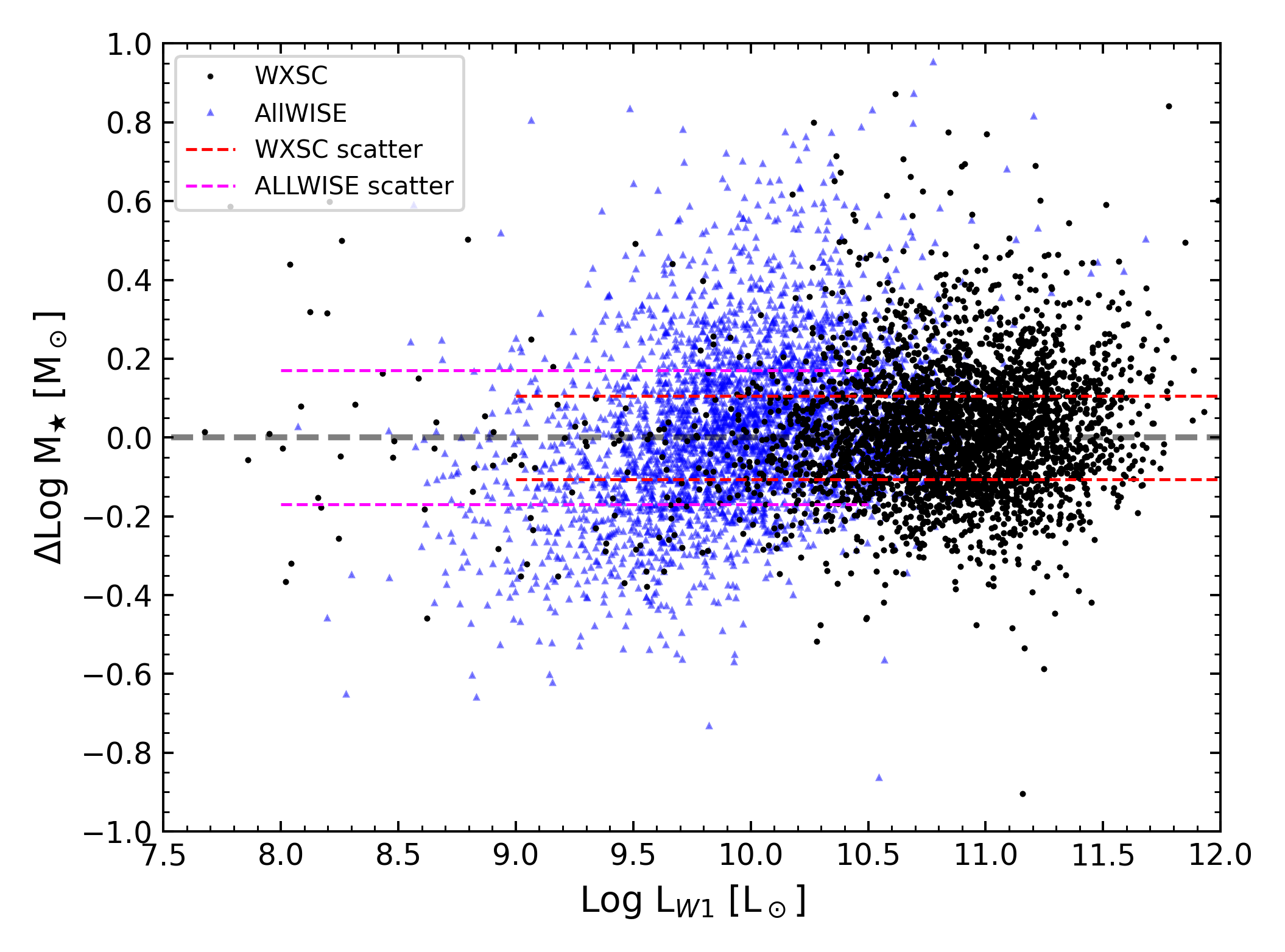}
\includegraphics[width=0.497\textwidth]{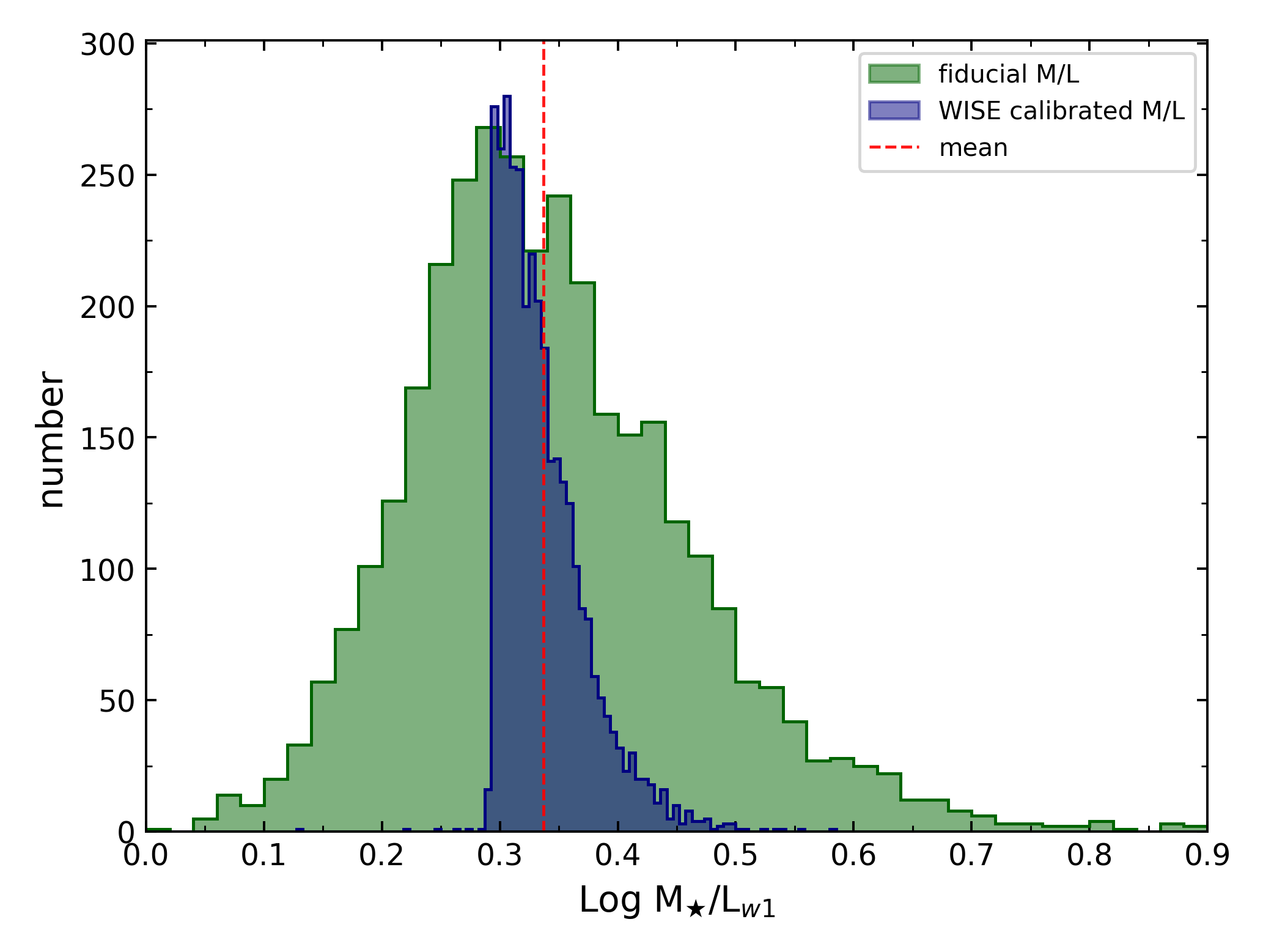}
\caption{\WISE\ \MIR\ stellar masses derived using only the \WISE\ W1 luminosity (Eq. 2).  Compare with the complete method results, Fig.~\ref{fig:fig5}, the scatter has increased (left) and the M/L is more restricted to the mean value $\sim$0.36.  It should be noted that although the random scatter is reasonable given the restricted M/L range,  they are in fact strongly correlated with M/L; i.e. different errors for galaxy types (e.g., early vs late-types).
}
\vspace{-10pt}
\label{fig:method1}
\end{center}
\end{figure*}

\begin{figure*}[!t]
\begin{center}
\includegraphics[width=0.497\textwidth]{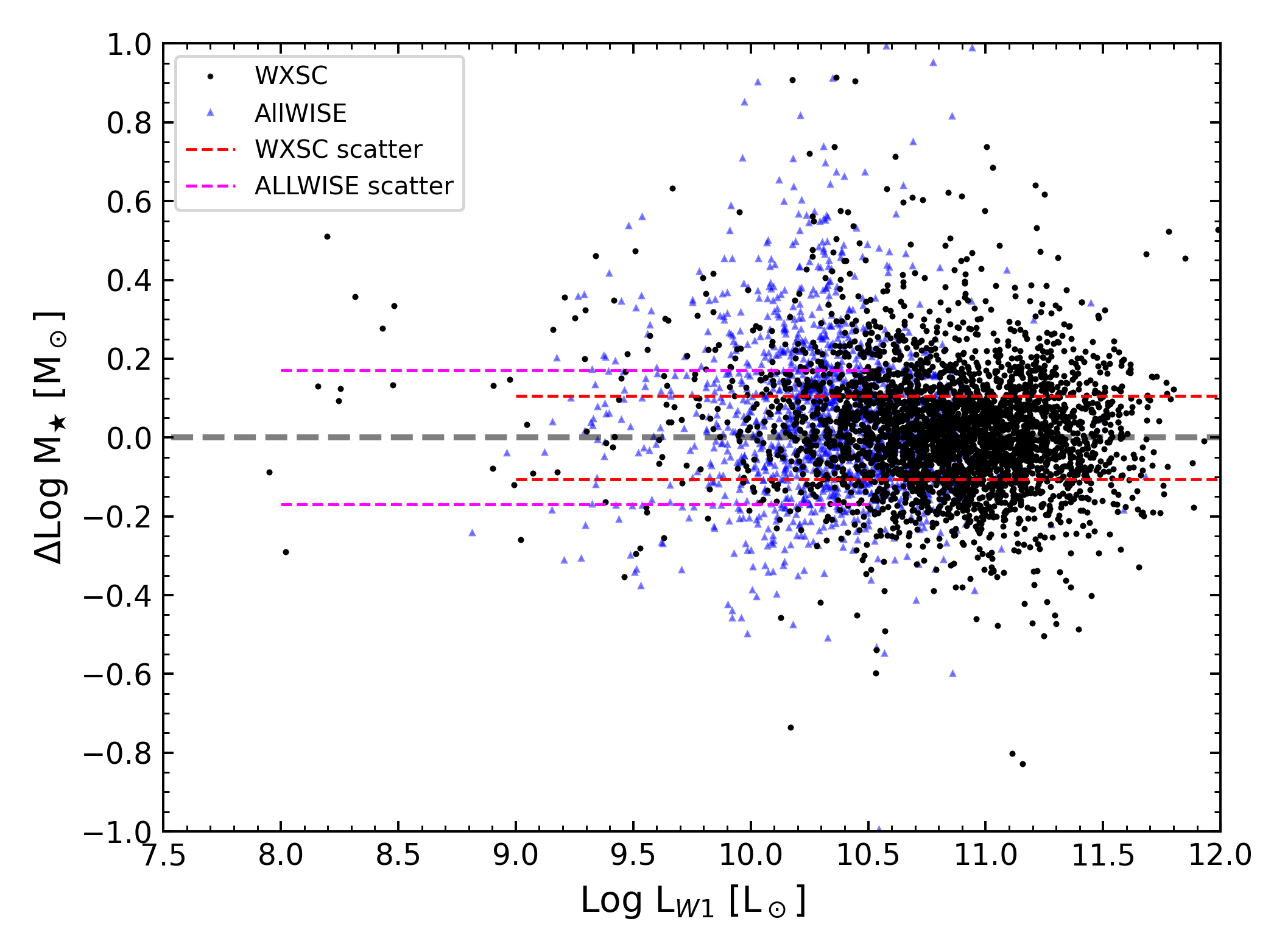}
\includegraphics[width=0.497\textwidth]{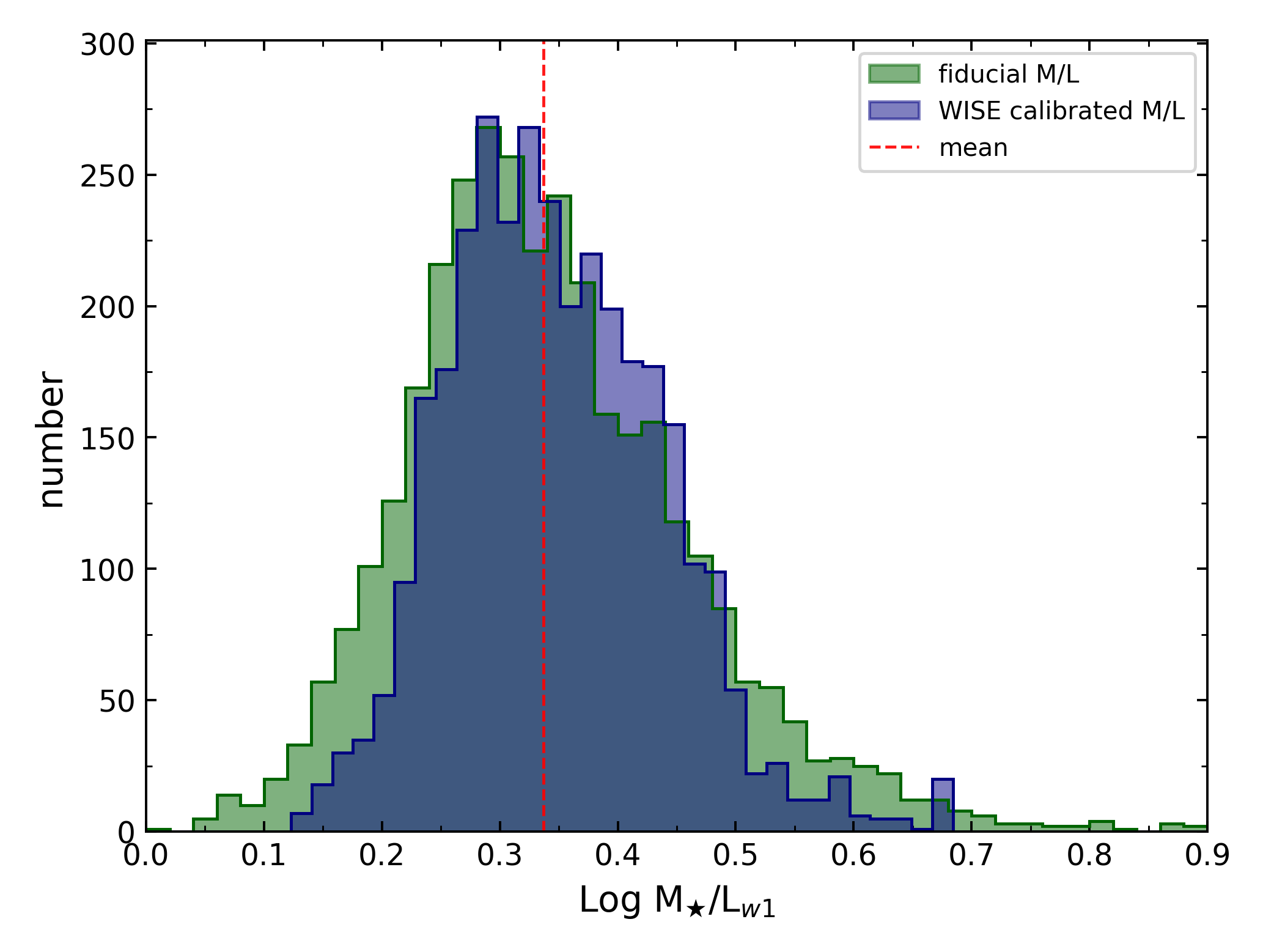}
\caption{\WISE\ \MIR\ stellar masses derived using only the \WISE\ \WIWII\ color (Eq. 3).  Compare with the complete method results, Fig.~\ref{fig:fig5}, there are fewer sources at low mass (left), while the M/L covers nearly all the range observed in the fiducial sample.
}
\vspace{-10pt}
\label{fig:method2}
\end{center}
\end{figure*}

Step one is to apply rest-frame correction to the measured fluxes.   As we show in the Appendix, the diversity of galaxy types and their associated variation in energy distributions means that it is non-trivial to carry out these corrections for the \MIR, requiring careful SED modeling.  We note, however, that the W1 3.4\micron\ correction is approximated by an exponential law, that works well for most galaxies and appears robust for redshifts $<$ 0.5 (see  Eq. A.1 and Table~\ref{table2} for the correction to the W1 flux).  Likewise, a simple polynomial (cubic) works well for the \WIWII\ color; see Eq. 2 and Table~\ref{table2}.  However, even in these instances, there will be some variation across galaxy types; graphically, these k-corrections are shown in Fig.~\ref{fig:kcorrections1} of the Appendix.  

For the ISM-sensitive color, \WIWIII\ (and \WIIIWIIII), there is no simple prescription, and we, therefore, do not recommend using a simplified function.  The k-corrections for these redward bands are graphically shown in Fig.~\ref{fig:kcorrections2}.   On the other hand, if a source has a clear W3 detection, it likely has dusty star-formation and you can be sure it will be a later type (Sb, Sc, etc).  You can use the curves in the Appendix with the appropriate correction and attempt a rest-frame correction to the W3 flux (in which case Eq. 4 may then be used).  

As shown in \cite{Cluver14} and in this current study, the rest-frame corrected \WISE\ color \WIWII\ is a reasonable and reliable tracer of the mass-to-light, and it alone (with the W1 luminosity) will provide a very effective stellar mass estimate.  We thus recommend that the simple 
relation given in Eq. 3 be used for this purpose, assuming a \WIWII\ color that has a S/N of at least 5.  For those cases where W2 is faint or undetected, resort to the relation given in Eq. 2, with an expected uncertainty of about 30\%.

In summary, to derive a stellar mass using a minimal set of information, 
we recommend the following:
\begin{enumerate}

\item Convert the observed W1 mag (total mags are best) to flux density (e.g., the zero point correction is 309.68 Jy for W1 Vega mags; see \cite{Jarrett2011}).  

\item Correct the W1 flux to the rest frame, you can look up the scale factor relevant to your source redshift from Fig.~\ref{fig:kcorrections1}, or apply the exponential law relation given in Eq. A1. and Table~\ref{table2}.  The rest flux is found by scaling by this factor; accordingly, 
f$_{\rm W1}$[rest] = $\frac{f_{\rm W1}[rest]}{f_{o}}\times$f$_{\rm W1}$[obs].

\item Compute the W1 in-band luminosity, \Lw, by first converting your rest-frame flux to absolute magnitude (use the distance modulus after converting your redshift to luminosity distance), and applying Eq. 1.

\item If you have a W2 detection and the resulting \WIWII\ uncertainty is better than 0.2 mag, you should attempt a k-correction using Fig.~\ref{fig:kcorrections1} or using the polynomial given in Eq. A2 and Table~\ref{table2}.  The rest-frame color is  [\WIWII]$_{rest}$ = [\WIWII]$_{obs} - \delta$[\WIWII].  Note the two redshift ranges for the polynomial coefficients specified in Table~\ref{table2}.  

\item  The stellar mass may be estimated from simply the \Lw\ using Eq. 2.  However if you have a good \WIWII\ color, we recommend instead using Eq. 3 (or a weighted combination of the two), in which case a better range in mass-to-light is covered.  The stellar mass is then:  \Lw $\times$ \Yw.

\end{enumerate}

This simplified stellar mass prescription is still very effective, producing accurate stellar masses, with only a few\% additional scatter compared to the comprehensive treatment.   This is shown in Figures \ref{fig:method1} and \ref{fig:method2}, where we show the stellar masses compared to the fiducial for the two simple cases: (1) using only the W1 luminosity, and (2) using only the \WIWII\ color.  
Although the random errors look reasonable, we note that errors are strongly correlated with M/L; i.e. different errors for galaxy types (e.g., early vs late-types).

In the first case, lower luminosity sources have estimated masses, but they also have a restricted range in M/L, between 0.3 to 0.42.  The total scatter has increased by 5-10\%. In the second case, there are far fewer sources with good quality \WIWII\ color, but for those that do have a color, the results are excellent, with only 2\% increased scatter compared to the comprehensive derivation, while the full range in \Yw\ is covered (almost identical to the fiducial sample used to calibrate the masses).  It is clearly advantageous to have color available to estimate accurate stellar masses.

\subsection{Example}
As an example of using these prescriptions, we consider 
WXSC-2dF source 96128; Ra, Dec(J2000) $=$ (358.70312, -33.04600 ).  
The SED is shown in 
Fig.~\ref{fig:SED}, which includes 2MASS and WXSC measurements, as well as the template used to restframe-correct the observed fluxes.   
The observed W1 mag is 13.200$\pm$0.015 (1.6222$\pm$0.0225 mJy), and colors(mag) \WIWII$=$ 0.178$\pm$0.027, 
\WIWIII$=$ 3.794$\pm$0.038, 
\WIIIWIIII$=$1.653$\pm$0.190.  The CMB-frame redshift is 0.06057 (helio is 0.06150), and the corresponding luminosity distance is 271.6 Mpc (co-moving, 256 Mpc), so the distance modulus is 37.17 mag.  The Log \Lw\ and \Mstar\ from the WXSC (which uses SED-fitting for k-corrections and the full-up \Mstar\ method) are, respectively, 10.811$\pm$0.028\,\Lsun\ and  10.272$\pm$0.079\,\Msun.  The M/L is thus 0.29.\\

\underline{K-correction:} 
For the redshift and using exponential function Eq A1. with coefficients (1.0 and -2.615; see Table~\ref{table2}), the scaling factor is 0.854.  That is to say, the W1 flux is brightened by 20\% due to the R-J tail shifting into the 3.4\micron\ band.  So we apply the correction to give a rest flux of 1.6222$\times$0.854 = 1.387 mJy.  Converting this flux back to magnitudes (13.37 mag), and applying the distance modulus, the resulting absolution magnitude, M$_{W1}$, is -23.80 mag.  And finally using Eq. 1, we arrive at the W1 luminosity, Log \Lw\ = 10.816\,\Lsun.  (note:  if you do not apply a k-correction to the flux, the Log luminosity would be $\sim$0.08\,dex too large).  From Eq. 2, the estimated Log\, \Mstar\ is 10.32\,\Msun, and the resulting \Yw\ is 0.32.  

Since the \WIWII\ color has a good quality S/N, we proceed with a rest-frame correction.
At the redshift of the source, the change in color is (Eq. A2, or look up with Fig.~\ref{fig:kcorrections1}) is 0.036 mag, but may also range from 0.02 to 0.06 mag depending on the galaxy type (see the figure).  
Using the average value of 0.036 mag,  the rest frame color is then $0.178 - 0.036 = 0.142$ mag.  It then follows from Eq. 3, that \Yw\ is 0.30.  Hence, the resulting Log\, \Mstar\ is 10.29\,\Msun. This mass is 7\% smaller than the estimate using only the luminosity because the mass-to-light is slightly smaller, appropriate to a star-forming galaxy (which it clearly is, see Fig.~\ref{fig:SED}.  It is also nearly equivalent to the full-up value, showing that even the simplest of prescriptions works very effectively.   
If no k-correction was applied to the color, the M/L is lower (0.27) and the stellar mass is correspondingly lower by 0.04 dex (10\%).


\section{Discussion}

Galaxies, enclosed within their dark-matter-dominated halos, are comprised of stars in all stages of their growth and evolution. The aggregate stellar mass will depend on the stellar content, which we observe and estimate through its light. Given the observed band or set of bands, and the initial mass function,  we can infer the mass distribution and hence, total mass from stars. It is clearly a complex problem for extragalactic sources given that only aggregate (population-combined) measurements are typically available, thus washing out any differences galaxies may have from one to another.  
Typically a single M/L is adopted for a given band for convenience (and standardization advantages), notably with large surveys where limited information is available.  However, it has been shown that adding some simple additional information -- notably the colors -- provides some stellar population information that informs a more diverse range in \Y\ and increases fidelity to the resulting \Mstar.  Adding additional information, such as star-formation history and morphological structure may further improve the M/L coverage.

Our new scaling relations, derived simply from the rest wavelength \WISE\ W1 (3.4\micron) total fluxes, and (\WIWII) and (\WIWIII) colors, provide \Mstar\ estimates for galaxies that range from dwarf to giant galaxies (10$^7$ - 10$^{11.5}$\,\Msun) with $\sim$25-40\% accuracy (relative to the fiducial masses used to create the relations) while covering a M/L range between 0.2 to 0.6 that reflects the dominant stellar populations in the individual galaxies.  The mean \Yw\ we compute is 0.34 $\pm$ 0.11, consistent with recent derivations using the mid-infrared 
\citep[e.g.][]{leroy+2019}, but somewhat lower than earlier work with Spitzer and \WISE\ that had values close to 0.5 to 0.8 
\citep[e.g.][yet for perspective, these larger \Y\ values are seemingly appropriate to early type, spheroidal galaxies]{meidt+2014}.

Our newly derived masses should be slightly lower than previous estimates since we are now using the new GAMA/KIDS/VIKING-G23 stellar masses derived from sophisticated SED fitting with opt-IR multi-band photometry, as well as our new, updated \WISE\ resolved galaxy photometry.
These masses should have a typical (normal galaxy) uncertainty of 25-30\% or 0.10-0.12 dex in the Log, but most importantly, be more appropriate
to those with significant population differences; e.g., early vs late-types; passive vs. active types, which can be significantly (2$\times$) different in M/L value
\citep[$cf.$][]{schombert+2022}.
Even the simpler prescriptions we provide to estimate stellar masses from only the W1 flux or the \WIWII\ color (see Section 5) have relatively good (30-50\%) mass estimates for all luminosity/mass ranges, and at least with the single color method,  cover a wide range in \Yw.
Since galaxy evolution work is focused on how galaxies grow and evolve in their home environment over cosmic time, thus covering the full range of masses that populate the field, groups, and clusters,  it is vital that an accurate M/L and aggregate stellar mass be estimated for the entire range, especially the most difficult to observe and study, the dwarf galaxy regime.  

These new relations should improve ongoing and future studies, provide a stronger `control' to cross-compare physical parameters and construct more representative SFR-\Mstar\ diagrams, the so-called `star formation main sequence' that is a powerful diagnostic tool of past-to-present SF history.  It is particularly important now to have this capability since next-generation large-area surveys are happening in this decade, notably from radio, optical and X-ray surveys, which all depend on having a stellar mass estimate for individual galaxies.   \WISE\ covered the entire sky, reaching depths of 20 to 30\,$\mu$Jy, and hence may provide the necessary information to estimate \Mstar.  If the galaxy is local, $z$ $<$ 0.1 to 0.2, then it is likely resolved in the 3.4\micron\ band, requiring a more careful extraction, typically not conducive to automated pipelines because of blending, contamination, and shredding issues that plague galaxy photometry.   

The \WISE\ resolved catalog, WXSC, was created to fill this need, to carefully measure local universe galaxies, and to extract basic properties that include colors and total fluxes. It was effectively used to derive scaling relations with \Mstar\ and SFR \citep[e.g.,][]{Cluver17,jarrett+2019}, and is being used to study local galaxy samples \citep[e.g.][]{ogle+2019,yao+2020,naluminsa+2021,bok+2020,bok+2022,healy+2021}. 
Since the WXSC is not fully complete for the local universe (as of this writing, it contains about 200,000 sources or an order of magnitude less than is required for the total volume),  it remains an ongoing task to build and hence estimate the stellar masses for galaxies in the local universe.

\section{Summary}


Updating from the original work in 
\cite{Cluver14},
we derive new \MIR\ based scaling relations that are used to estimate global stellar masses for galaxies.
The primary photometric data are drawn from the \WISE\ Extended Source Catalogue (WXSC), developed to extract and characterize galaxies that are resolved in the W1 3.4\micron\ band.  Compact and unresolved galaxies are drawn from the archival ALLWISE catalog, with further corrections to these measurements to capture the total flux of compact-resolved sources.
The stellar masses, used as the fiducial calibration set, come from the new
 DR4 Catalogue of the 
the GAMA-KiDS-VIKING survey of the southern hemisphere G23 field.  

Restricting to 
redshifts $<$ 0.15 and focusing on normal galaxies that span 6 orders of magnitude in
W1 luminosity, we construct 
three scaling relations:  the \Lw\ versus the \Mstar, and the \WISE\ \WIWII, \WIWIII\ colors versus the
3.4\micron\ mass-to-light (\Yw).  The color relations are 
sensitive to a variety of galaxy types from early-type or passive to late-type or star-forming.  
Combining the scaling relations using inverse-variance weighting, we produce a mean stellar mass estimate that has a residual scatter (relative to the fiducial GAMA stellar masses) 
 better than 0.10-0.12 dex accuracy for most galaxies (with W1 luminosities $>10^{9}$\Msun) and 0.15-0.25 dex for lower mass dwarf galaxies.  These new scaling relations will enable stellar mass estimates for any galaxy in the sky detected by \WISE\ with $\sim$25-50\% fidelity.\\
 
Detailed highlights include the following:
\begin{itemize}

\item The DR4 G23 catalog provides 16,000 high-quality stellar masses that are
used as the fiducial calibration set.   Cross-matching with the \WISE\ photometry catalog returns 3374
and 3297 resolved and compact sources, respectively, with S/N $>$ 5.   A slighter smaller fraction of resolved sources have a full set of \WISE\ colors.

\item  Essentially the Log \Lw\ scales linearly with Log \Mstar, although we fit a stiff (cubic) polynomial that accounts for a small 
 upturn at the high-mass end (see Eq. 2).   The roughly linear relation means that a single or tight-range in mass-to-light adequately ($\sim$30-50\%) describes the stellar mass in the W1 band.   We find a mean \Yw\ value of 0.35 $\pm$ 0.11.  Relative to the fiducial mass, the derived mass has a statistical scatter of about 0.12 dex, but several \% higher for the low mass regime.  It should be noted, however, embedded are  
 strong differential systematics for early and late-types -- which can amount to as much as 0.6 dex (peak to trough) in log mass.  These systematics cannot be meaningfully encapsulated with simple Gaussian statistics.
 
 \item Unlike the single luminosity scaling relation, 
 colors trace a more representative range in M/L.  
 The \WISE\ \WIWII\ color scales linearly with the Log \Mstar, see Eq. 3.  It covers the expected range that normal galaxies, from passive to dusty star-forming, exhibit, with a resulting \Yw\  range of 0.7 to 0.16, respectively.  It has a 25-30\%  residual scatter.
 
 \item Color \WIWIII\ has a more complex trending with Log \Yw, steepening from linearity at colors redder than 3.1 mag.   The flat \Yw\ regime and roll-off to much lower values  
 conveniently fit with a Schechter Function, arriving at a scaling relation (Eq. 4) that
 has a residual scatter from 23-38\%, the higher variance at the red color end where the mass-to-light is changing more rapidly and less constrained.
 
 \item Using inverse-variance weighting, we create a mean-combined stellar mass from the three scaling relations. In order to cover a wider range in mass-to-light and create an optimal, maximally informative combination, we de-weight the first scaling relation to emphasize the power of the \MIR\ colors to better sample the variety of real galaxies.  The resulting scatter in the Log \Mstar\ is $\sim$0.11 (in Log dex) for most galaxies, 10$^{9}$ to 10$^{12}$ \Msun\ in W1 luminosity, rising to 0.16-0.17 dex for those with faint flux levels and lacking full-color information.

 \item In terms of mass-to-light, the derived (weight-combined) \WISE\ \Yw\ has a mean value of 0.34 and a FWHM width of 0.21 (insofar as a Gaussian can be modeled), while the extreme wings of the distribution reach values as low as 0.2 and as high as 0.6.
 
 \item We also provide simpler prescriptions for estimating the mass, using (1) only the rest-frame W1 luminosity, or (2) only the rest-frame \WIWII\ color. In conjunction, we provide a (far) simpler prescription for estimating the rest-frame correction to the W1 flux and the colors (details in the Appendix).   These simple prescriptions produce satisfying results, with with a $\sim$few\% more scatter compared to the full-up method, while also (using color) covering the full range in \Yw.

\end{itemize}

\begin{acknowledgments}
T.H.J. acknowledges support from the National Research Foundation (South Africa).
M.C. is a recipient of an Australian Research Council Future Fellowship (project No. FT170100273) funded by the Australian Government. 
This publication makes use of data products from the Wide-field Infrared Survey Explorer, which is a joint project of the University of California, Los Angeles, and the Jet Propulsion Laboratory/California Institute of Technology, funded by the National Aeronautics and Space Administration.
GAMA is a joint European-Australasian project based around a spectroscopic campaign using the Anglo-Australian Telescope. The GAMA input catalog is based on data taken from the Sloan Digital Sky Survey and the UKIRT Infrared Deep Sky Survey. Complementary imaging of the GAMA regions is being obtained by a number of independent survey programs including GALEX MIS, VST KIDS, VISTA VIKING, \WISE, Herschel-ATLAS, GMRT, and ASKAP providing UV to radio coverage. GAMA is funded by the STFC (UK), the ARC (Australia), the AAO, and the participating institutions. The GAMA website is located at http://www.gama- survey.org/. 

\end{acknowledgments}

\vspace{5mm}
\facilities{\WISE, AAT}


\appendix
\section{\WISE\ Rest-frame Corrections}   

Unlike the near-infrared, rest-frames corrections are non-trivial for the \WISE\ \MIR\ bands due to strong spectral features, including the [Si] absorption at 10\micron, broad and prominent PAH emission bands at 6, 8, 11.3\micron\ and [NeII] emission at 12.8\micron, in addition to the underlying stellar continuum and warm dust continuum.  In general, simple power laws are limited to describing the shift in flux with (1+z) photometric band-shifting.    
From one galaxy to the next, there are varying  
 contributions from the stellar continuum (current and past history), emission lines, and ISM components (e.g. dust) that arise from current star formation activity, rendering a composite spectrum that reflects the diversity of galaxies observed in the local universe.   
 
 The approach adopted by the WXSC team is to model the spectrum using composite (pre-constructed) templates that range from the blue (optical) to the mid-infrared ($<$40\micron) and measure the synthetic fluxes (integrated across the photometric bandpasses), comparing them to the observed photometry.  
  We identify the best fit template, scaling the ``observed" measurements to the equivalent ``rest" measurements.  The list of composite templates we deploy are from  \citet[]{Brown2014} and Spitzer-SWIRE/GRASIL \cite[]{Polletta2006,Polletta2007,Silva1998},
 totaling over 135 composite templates from a wide range of morphological and activity types.
 Further details of the SED template fitting are given in \citet[]{Jarrett13,Jarrett2017}.

\begin{figure*}[!thb]
\begin{center}

\includegraphics[width=0.49\textwidth]{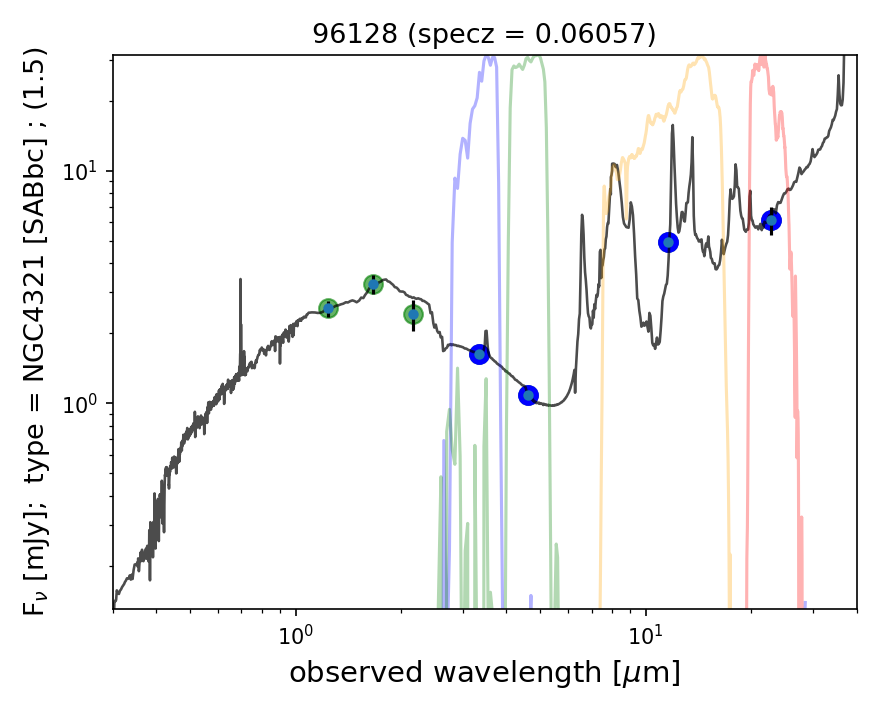}
\includegraphics[width=0.49\textwidth]{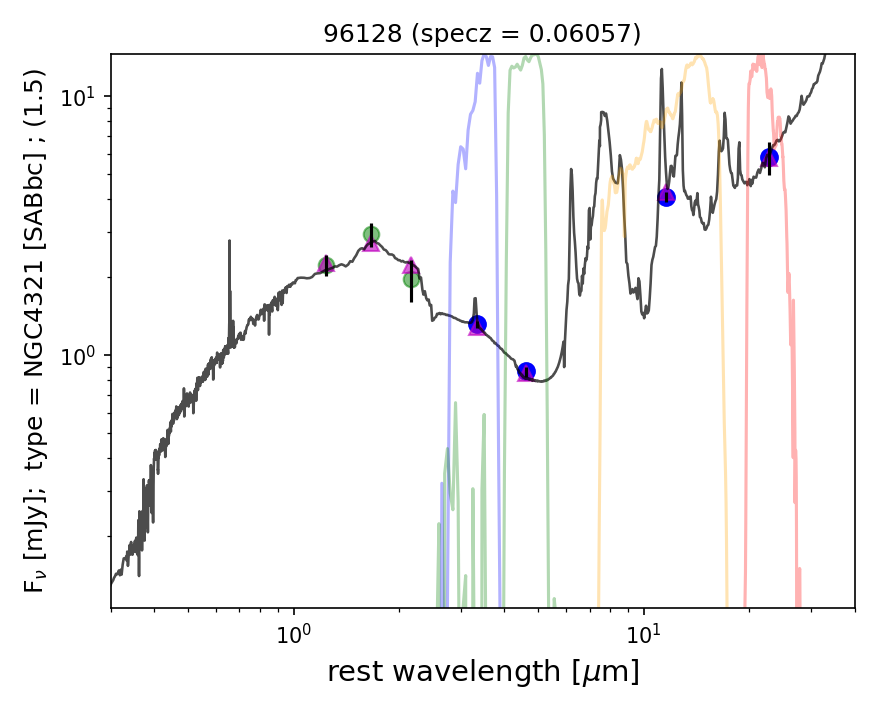}
\vspace{-10pt}
\caption{Spectral energy distribution (SED) of the WXSC-2dF source 96128, whose redshift is 0.0606.   The near-infrared data (green points) are total fluxes from the 2MASS Extended Source Catalog, and the \MIR\ (blue points) is from the \WISE\ XSC.
The best fit (black line) was found to be that of the NGC\,4321 template \cite[from][]{Brown2014}, which has a morphological type of SABbc.
LEFT panel shows the observed (i.e., redshifted) fluxes, and RIGHT panel the k-corrected rest-frame fluxes (note the pink triangles are the expected fluxes based on synthetic photometry).  Also shown are the \WISE\ filter traces (scaled to unity; blue, green, orange, and red lines), as given in \citet[]{Jarrett2011}.  
}
\label{fig:SED}
\end{center}
\vspace{-15pt}
\end{figure*}

 An example SED is shown in Fig.~\ref{fig:SED}, a typical spiral/disk galaxy from the GAMA/KiDS-s sample in the WXSC.  It shows the observed fluxes, as measured by 2MASS and \WISE\ (left panel), with the best fit spectral template indicated by the solid line.  It features the well-known stellar "bump" that peaks in the H-band (1.6\micron), characteristic of the luminous evolved population.  The R-J tail of the continuum extends into the \MIR\ where the \WISE\ W1 3.4\micron\ and W2\micron\ bands pick up the emission. It then abruptly transitions to ISM emission, warm dust continuum which now rises with longer wavelength, PAH and [NeII] emission in the broad W3\,12\micron\ band, and finally steep rising dust continuum in the W4 23\micron\ band.   Correcting to the rest frame using the synthetic photometry of the spectral template, the resulting rest-frame SED is shown in Fig.~\ref{fig:SED}, right panel.   
 
 When the rest fluxes land on the expected synthetic fluxes (pink triangles), the fit quality should be good, as indicated by the $\chi^2$ metric, as is the case in the example given here.  There is a finite set of templates, and the photometric quality determines fit quality, and hence not all fits are high quality, notably for those with only a few points (i.e., faint) or with extreme
 properties that are not well characterized by our suite of templates.  For the most part, the k-correction imparts less than 5-10\% uncertainty for most sources that have redshifts $<$ 0.3 (see also the Appendix in 
 \citet[]{yao+2022}).

 A cautionary note about rest-corrections:  as pointed out by \cite{hogg+2002}, there is an additional (1+z) scaling that must be applied when using spectral luminosities ($\nu$L$_\nu$).  For example, star formation rates are computed using spectral luminosities \cite[e.g.,][]{Cluver17} and hence would have the luminosity normalized by (1+z).  Whereas for bolometric luminosities, this factor does not apply, which is the case for the stellar masses in which the W1 in-band luminosity (see Eq. 1) is used for the \Yw\ scaling relations. 
 
 \vspace{10pt}
 
To show the complexity of the rest frame corrections, as a function of redshift, we plot the expected correction of the W1 flux (mJy units) and the \WISE\ colors (in Vega mag units) for a set of templates that range from early to late types (bulgy to disk).  
These are normal galaxies and their expected optical-infrared emission spectra, from which we compute synthetic fluxes, comparing the observed with the rest flux and colors.  (note, we are not showing the exotica, which may look completely different; e.g. QSOs).  

\begin{figure*}[!thb]
\begin{center}
\includegraphics[width=0.49\textwidth]{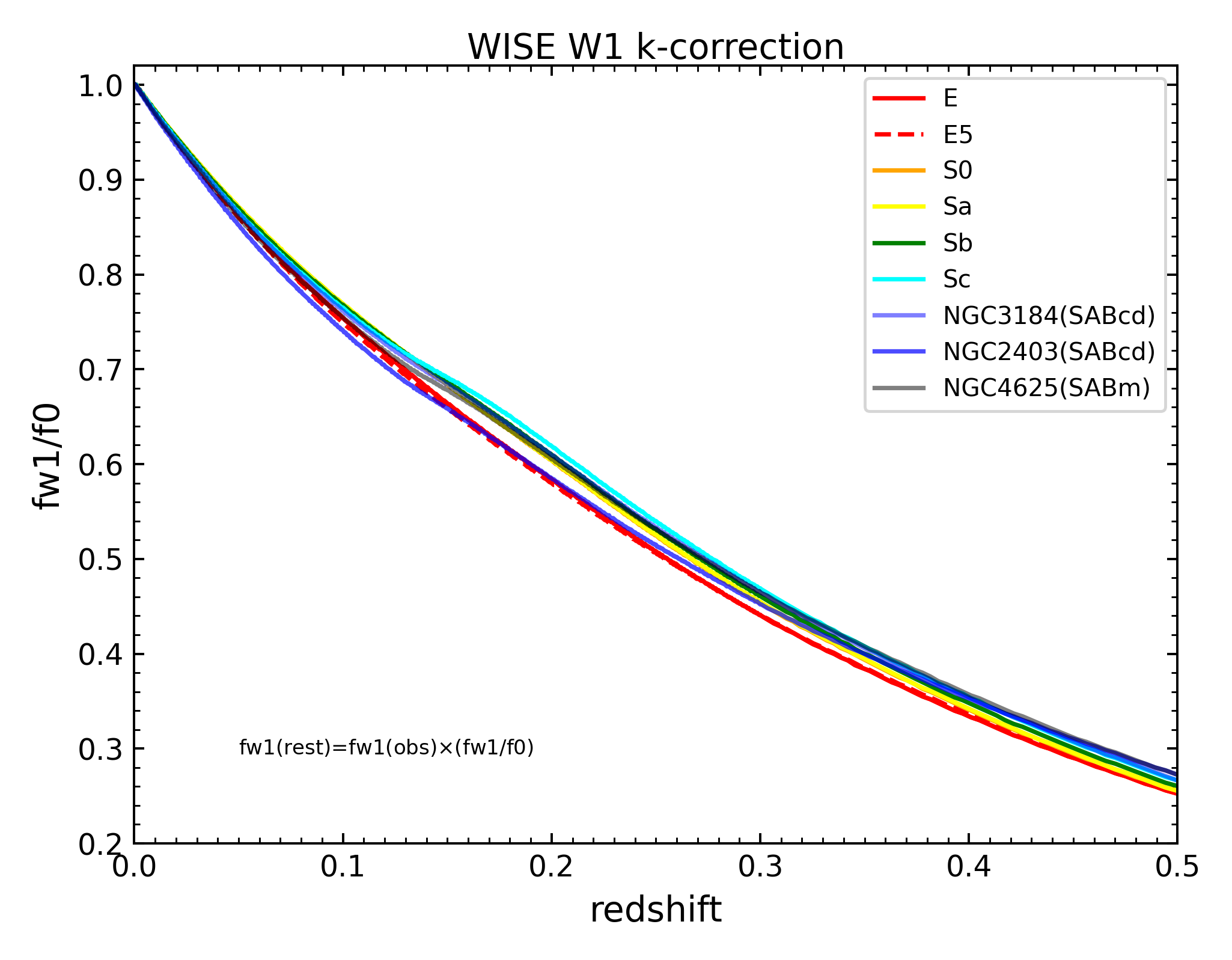}
\includegraphics[width=0.49\textwidth]{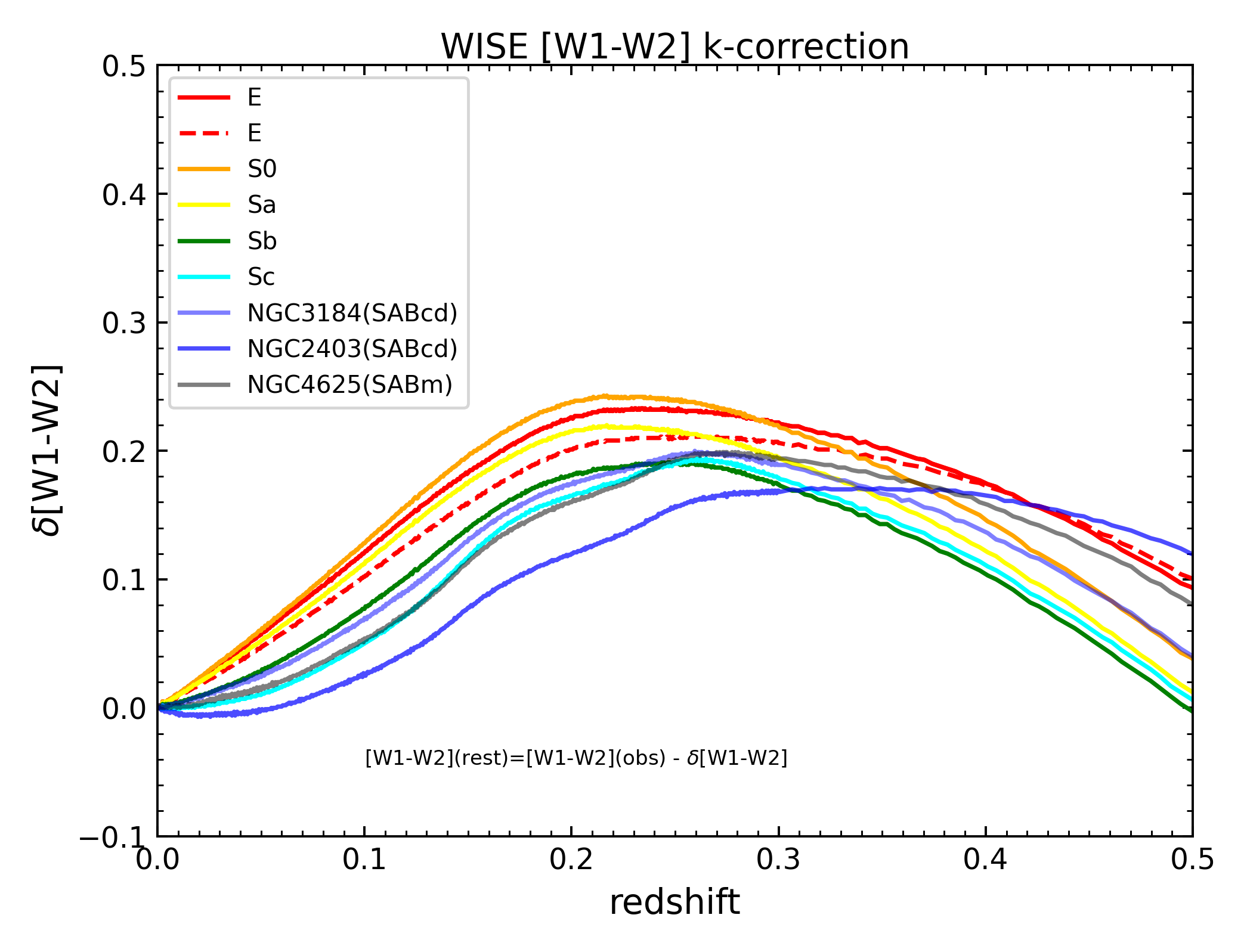}
\vspace{-10pt}
\caption{Rest-frame corrections for \WISE\ W1 3.4\micron\ (left) and \WIWII\ (right) based on SED composite templates that cover the galaxy types from early-to-late (bulge-to-disk dominated). The W1 scaling correction is approximated by an exponential function, see Eq. A1, and the color by a cubic polynomial (Eq. A2);  both based on the Sb template.
}
\label{fig:kcorrections1}
\end{center}
\vspace{-1pt}
\end{figure*}

Similar to the near-infrared K-band band (at 2\micron), the \WISE\ W1 3.4\micron\ band gets brighter with (1+z) because of the `peak' stellar bump at 1.6\micron\ shifts longward toward the K and W1 bands, increasing their flux (relative to rest, ignoring the 4$\pi$r$^{2}$ flux dilution).   As shown in Fig.~\ref{fig:kcorrections1}, upper left panel, the expected change in flux with redshift is well-behaved, an exponential function (or power-law), similar for all galaxy types.  This demonstrates how most normal galaxies have the same near-infrared (1-4\micron) spectrum (i.e., the stellar bump from evolved, luminous K/M stars), nearly a "standard candle" which may be exploited in photometric-redshift studies.   The k-correction is the scale factor shown on the Y-axis, it is less than unity to reduce the observed flux to the rest value.  Do note that the correction is significant, 25\% at z=0.1, and nearly 75\% at z = 0.5.  By doing no correction and just assuming the fluxes and colors are 'rest-like', the ``rest" flux will be highly over-estimated, badly skewing results.  It is important to apply a k-correction to the flux (and colors, see below).  Below we provide simpler descriptions which, even if limited, are better than doing nothing. 

For convenience to the \WISE\ user, 
we fit an exponential (a power-law also works well) to the "Sb" template,
it is a good middle-ground, straddling all galaxy types;  it has the form:

 \begin{equation} \label{eq4}
 \frac{f_{\rm W1}[rest]}{f_{o}} = a_{0}e^{\alpha z}
 \end{equation}
where we fit for the scaling $a_{0}$ and index $\alpha$. The rest flux follows by scaling by this ratio; accordingly, 
f$_{\rm W1}$[rest] = $\frac{f_{\rm W1}[rest]}{f_{o}}\times$f$_{\rm W1}$[obs].
We find that the redshift range between 0 to 0.5 is well fit.
Beyond that redshift limit, it is best to do proper SED modeling to derive the k-corrections.  We emphasize an exponential or power-law is just an approximation of the rest-frame correction required for the W1 flux.
In Table~\ref{table2}, we provide the best fit coefficients for this limited redshift range.

Now consider the other bands or colors.  
The R-J stellar tail is still well-detected in the W2 at 4.5\micron.  Hence we expect the W2 correction to be simple as well.  However, the W2 band is subject to warm dust from the ISM and from AGN, it can be enhanced in SF galaxies and active galaxies; e.g., see Fig.~\ref{fig:fig1}, right panel.  The resulting W1-W2 color k-correction has more variation in galaxy type; see Fig.~\ref{fig:kcorrections1}, upper right panel.   Notably, the early types, E's n the diagram, are somewhat different from the disk galaxies, S types.  At higher redshifts, z $>$ 0.2, there is some mixing of types.   At this point, it is clear that more sophisticated modeling is necessary to estimate the rest-frame correction for a given galaxy.   However, in the spirit of deriving a simple and convenient relation that may be used for a large number of cases (i.e., normal disks in the local universe), we fit a 3rd order polynomial to the "Sb" template (again, a good middle ground, at least for z $<$ 0.2);  it has the form:

 \begin{equation} \label{eq4}
 \delta[W1-W2] = a_0 + a_{1}z + a_{2}z^{2} + a_{3}z^{3}
 \end{equation}
where we fit for the a-coefficients.  The rest-frame color is simply:  [\WIWII]$_{rest}$ = [\WIWII]$_{obs} - \delta$[\WIWII]. Here we fit two redshift ranges, 0-0.25 and 0.25 - 0.5 to accommodate the non-linear curve.  The coefficients are specified in Table~\ref{table2}.   Once again, this is only an approximation, notably for disk galaxies.

\begin{figure*}[!thb]
\begin{center}
\includegraphics[width=0.49\textwidth]{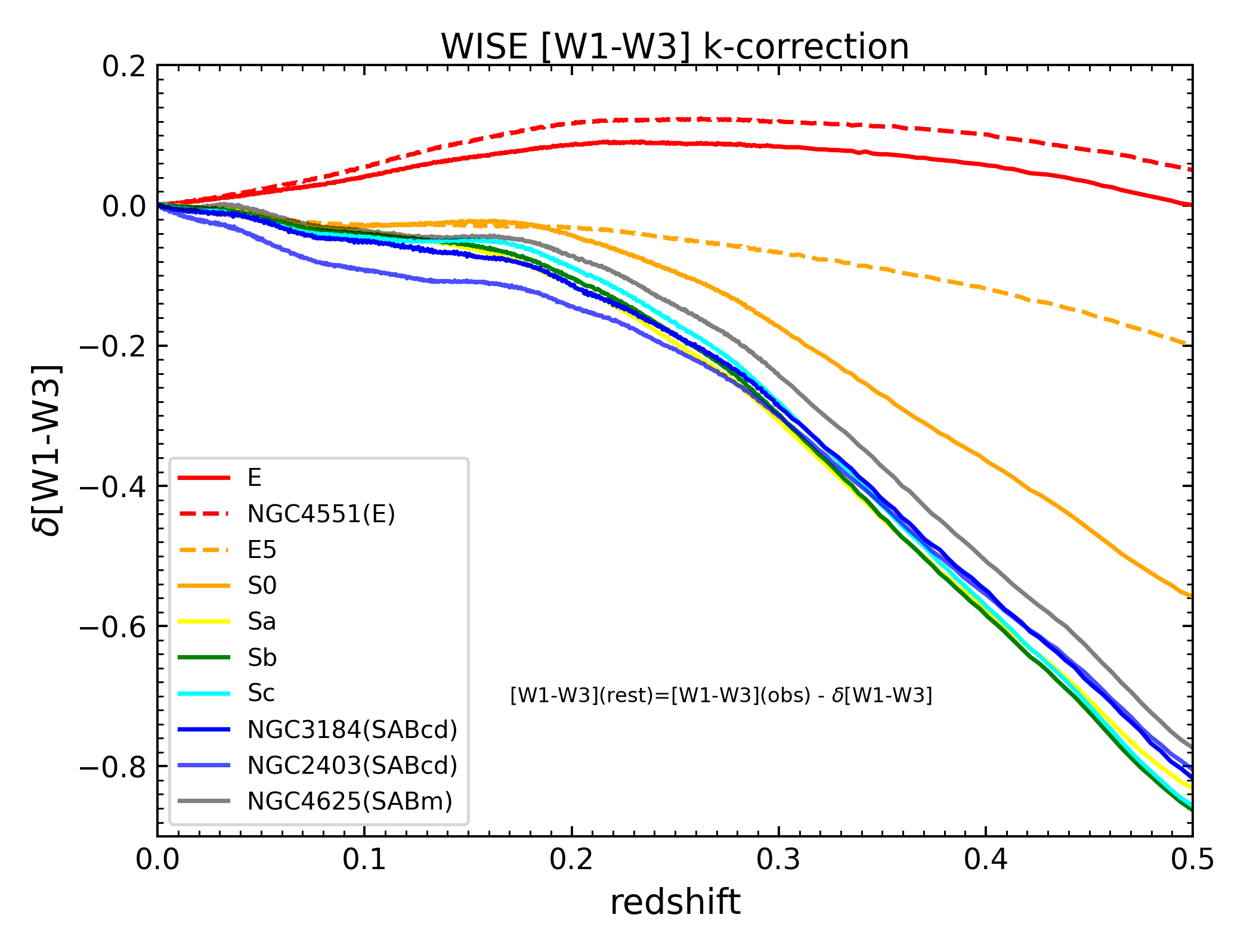}
\includegraphics[width=0.49\textwidth]{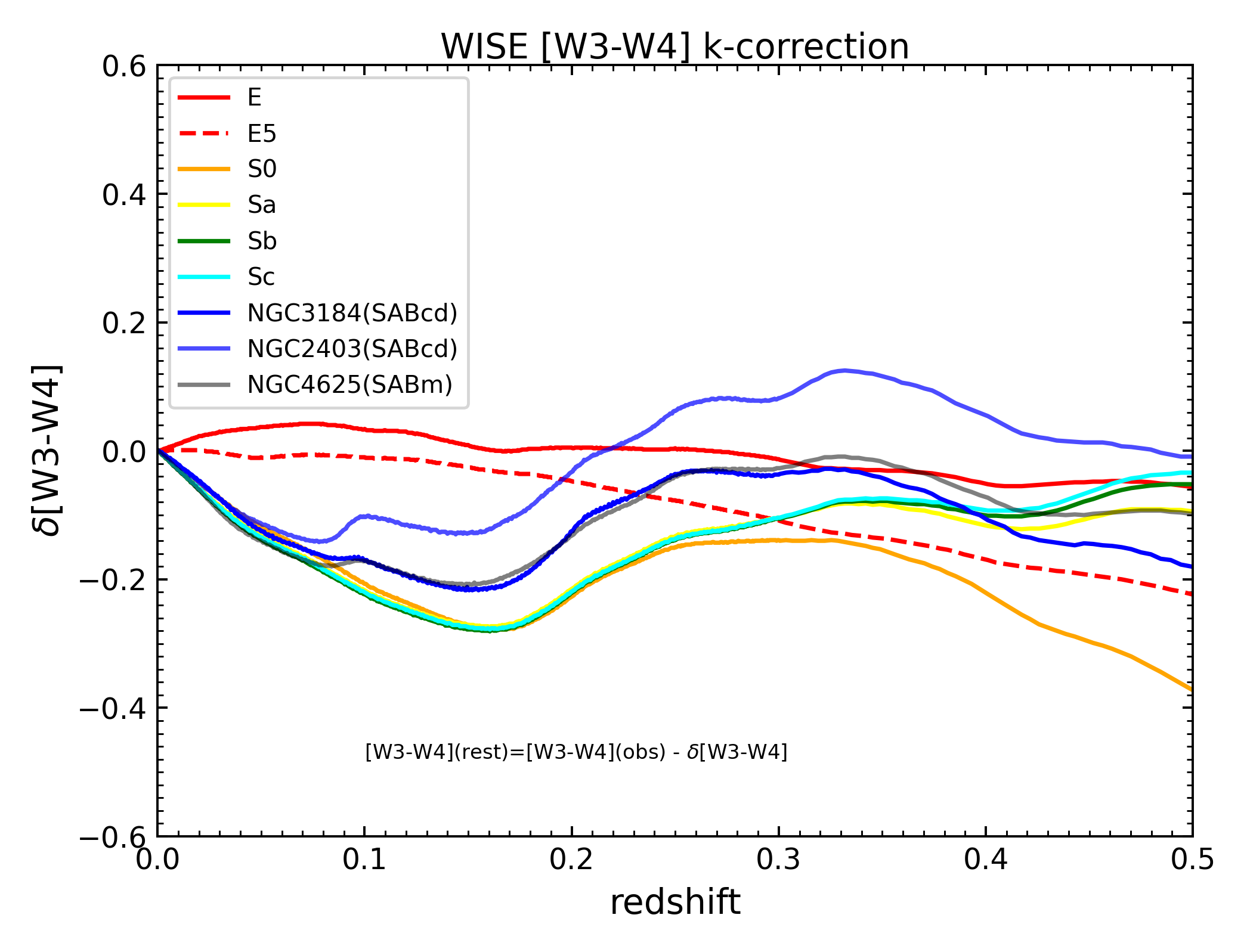}
\vspace{-10pt}
\caption{Rest-frame corrections for \WISE\ \WIWIII\ color (left) and \WIIIWIIII color (right) based on SED composite templates that cover the galaxy types from early to late.}
\label{fig:kcorrections2}
\end{center}
\vspace{-1pt}
\end{figure*}

Finally, we consider the longward bands of \WISE.  Fig.~\ref{fig:kcorrections2} shows the expected k-corrections for the \WIWIII\ color (left panel), and the \WIIIWIIII\ color (right panel).   As demonstrated in Fig.~\ref{fig:SED}, these bands are subject to a number of absorption, emission, and continuum components that are unique to the galaxy type.  At low redshifts, z $<$ 0.1, there are two paths (E types, and S types) that may be approximated with an exponential function, for higher redshifts, there is no common ground to form a simple k-correction relation.   It is best to use SED modeling to derive the needed corrections or use these figures as a lookup table.

\begin{table*}[!h]
 \caption{k-correction Coefficients for Eq. A1 and A2} \label{table2}
\centering
\small{
\begin{tabular}{l c c c c }
\hline  
\multicolumn{4}{c} {$a_{0}e^{\alpha z}$  }\\
 W1 flux    &   a$_0$         &   $\alpha$ &  &    \\
 \hline

 z $<$ 0.5         & 1.0 & -2.614 &  & \\ 

\hline  
\multicolumn{5}{c} {$a_0 + a_{1}z + a_{2}z^{2} + a_{3}z^{3}$  }\\
\WIWII    &   a$_0$         &   a$_1$ &  a$_2$ & a$_3$ \\
 \hline
    
 z $<$ 0.25  & 0.000 & 0.142 & 9.193 & -26.980 \\ 
 
 0.25 $<$ z $<$ 0.5 & -0.040 & 2.258 & -6.306 & 3.875 \\ 

\end{tabular}
}
\end{table*}

\bibliography{WISE_calibration_revisited}{}
\bibliographystyle{aasjournal}



\end{document}